\documentclass{pj}
\usepackage{graphicx}
\usepackage{psfrag} 
\usepackage{amssymb} 
\usepackage{amsmath}
\usepackage{xcolor}

\newcommand{\mb}[1]{\mathbf{#1}}

\renewcommand\Re{\operatorname{Re}}
\renewcommand\Im{\operatorname{Im}}

\begin{document}
\setcounter{page}{1}
\pjheader{}

\title[]{Numerical comparative study between regularized Gauss-Newton and Conjugate-Gradient methods in the context of microwave tomography}
\footnote{\hskip-0.12in*\, Corresponding author:~Slimane~Arhab (slimane.arhab@univ-avignon.fr).}
\footnote{\hskip-0.12in\textsuperscript{1} Avignon Universit\'e, UMR 1114 EMMAH, 84018 Avignon Cedex, France.}

\author{Slimane~Arhab\textsuperscript{*, 1}}

\runningauthor{}


\begin{abstract}
The reconstruction of relative permittivity and conductivity in microwave tomography is carried out using regularized Gauss-Newton and Conjugate-Gradient iterative schemes. These two approaches are numerically tested and compared in terms of resolution, speed of convergence and robustness to noise. The numerical results show that for noiseless data, the regularized Gauss-Newton iterative scheme has a better resolution and a higher convergence rate, while the Conjugate-Gradient iterative scheme remains the fastest in terms of calculation time per iteration. For noisy data, both approaches give almost the same results, making the Conjugate-Gradient approach the most suitable for inverting experimental data for its autonomy and ease of implementation.
 
\end{abstract}

\setlength {\abovedisplayskip} {6pt plus 3.0pt minus 4.0pt}
\setlength {\belowdisplayskip} {6pt plus 3.0pt minus 4.0pt}

\

\section{Introduction}
\label{Introduction}

The purpose of microwave tomography is to reconstruct the complex permittivity-contrast of an object from the measurement of the scattered electromagnetic field. One of the most studied physical configurations consists of an incident wave radiated by an emitter that interacts with the object, which radiates a scattered field that we detect on receivers. This study configuration has applications in various research fields, without being exhaustive we can mention brain imaging \cite{scapaticci2012feasibility,bonazzoli2018parallel}, breast cancer detection \cite{fear2002confocal,li2004microwave,souvorov2000two,gharsalli2014inverse,klemm2009radar} or the characterization of water content in soil cores \cite{zhang2010qualitative,zhang2010microwave}. These situations raise an inverse problem of reconstruction of inclusions introduced into an object, of known complex permittivity-contrast, which we propose to study in this work. Here, the geometric dimensions of the object under study are comparable to the wavelength of the incident field, the complex permittivity-contrast considered has significant values. The interaction between the object and the incident wave is therefore that of a resonant regime that is the site of multiple interactions \cite{martin2006multiple,foldy1945multiple}. Such phenomena are not taken into account by Born's or Rytov's approximations for which the inverse problem is linear. For instance, in ground penetrating radar these approximations lead to reconstructions of the permittivity map with a limited resolution \cite{hansen2000inversion,meincke2001linear,deming1997diffraction}. \\

In this work, microwave imaging is therefore treated as a nonlinear inverse scattering problem, its resolution requires the construction of a forward model that rigorously describes the interaction between the object and the incident field radiated by the source-emitter \cite{lesselier1991buried,richmond1965scattering,sun2009novel,tsang2004scatteringnum,tsang2004scatteringthe}. This forward model is used to generate the reference data, namely, the field scattered on the receivers by the object with its inclusions, which we refer to as the actual object or the unknown of the inverse problem. This forward model is also used to calculate the descent directions of the Conjugate-Gradient \cite{bertsekas1995} and the regularized Gauss-Newton \cite{bertsekas1995}, both of which are iterative inversion schemes used in this work. The initial guess consists of the object without its inclusions that we assume to be known, the complex permittivity-contrast is then gradually reconstructed over iterations, at any point within the object under test. The descent directions update the complex permittivity-contrast in the sense of minimizing the cost functional, which expresses the discrepancy between the reference data and the simulated data for the estimated object at each step of the iterative process. The regularized Gauss-Newton direction corresponds to the solution giving the minimum of the regularized Gauss-Newton model, which is a regularized quadratic model used to locally approximate the cost functional, in this model is introduced an additive constraint term weighted by a Tikhonov parameter, which is updated during iterations. We use the Polak-Ribi\`ere formula to calculate from the gradient and the Hessian the direction of the Conjugate-Gradient, the gradient and the Hessian being those appearing in the Gauss-Newton quadratic model. The regularized Gauss-Newton direction once calculated is sufficient to iteratively reconstruct the complex permittivity-contrast of the object, while the direction of the Conjugate-Gradient must be weighted by a scale factor that we calculate analytically based on the Gauss-Newton quadratic model. The gradient and Hessian are calculated from the Jacobian, an operator that links the variation in complex permittivity-contrast in the object domain to the variation in the scattered field data calculated at the receivers. The expression of this Jacobian is obtained by the reciprocity gap functional method \cite{arhab2018high,roger1982reciprocity}, which is a formulation equivalent to the adjoint method \cite{norton1999iterative}. \\

The work we propose aims to numerically compare the regularized Gauss-Newton and the Conjugate-Gradient iterative schemes, on their performance in inverting noiseless and noisy data. The purpose is to show which of the two methods is the most suitable for inverting experimental data, based on criteria such as resolution, robustness to noise and ease of implementation. After this introduction, we describe in the second section the study configuration and the volume integral formulation used to build the forward model. In section three, we introduce the cost functional and formulate the inverse problem. In section four, we introduce the Gauss-Newton quadratic model, we calculate the regularized Gauss-Newton descent direction after modifying the Gauss-Newton quadratic model, by adding a constraint weighted by a Tikhonov parameter. We then calculate the Conjugate-Gradient descent direction and show how to analytically calculate the scale parameter that weights it. In part five, which is devoted to numerical results, we begin by introducing and setting all the numerical parameters involved in data modeling and reconstruction of the actual object. Then, based on a numerical observation, we give a practical way to choose at each iteration the value of the Tikhonov parameter, by extracting its values from the spectrum of the eigenvalues of the Hessian of the Gauss-Newton model calculated in the vicinity of the initial guess of the object. Inversion results on noiseless and noisy data by the two iterative approaches are then presented and compared with each other. Finally, a conclusion is given in section six. 

\section{Study Configuration and Volume Integral Formulation}
\label{sec:Study Configuration and Volume Integral Formulation}
\subsection{Study Configuration}
\label{subsec:Study Configuration}

\begin{figure}[h]
\centering
\begin{psfrags}
\psfrag{1}[][]{\scalebox{0.9}{${\rm J}_{_{l=1}}$}}
\psfrag{2}[][]{\scalebox{0.9}{${\rm E}^{^{\scalebox{0.7}{\rm s}}}_{_{m=2}}$}}
\psfrag{3}[][]{\scalebox{0.9}{${\rm E}^{^{\scalebox{0.7}{\rm s}}}_{_{m=3}}$}}
\psfrag{4}[][]{\scalebox{0.9}{${\rm E}^{^{\scalebox{0.7}{\rm s}}}_{_{m=M}}$}}
\psfrag{5}[][]{\scalebox{0.9}{${\rm r_{_d}}$}}
\psfrag{6}[][]{\scalebox{0.9}{${\rm R_{_d}}$}}
\psfrag{7}[][]{\scalebox{0.8}{$\begin{array}{l}(\varepsilon,\,\mu_{\scalebox{0.55}{0}})\\\chi=\frac{\varepsilon}{\varepsilon_{\scalebox{0.55}{0}}}-1\end{array}$}}
\psfrag{8}[][]{\scalebox{0.8}{${\mb x}_{\scalebox{0.55}{1}}$}}
\psfrag{9}[][]{\scalebox{0.8}{${\mb x}_{\scalebox{0.55}{3}}$}}
\psfrag{10}[][]{\scalebox{0.8}{${\mb x}_{\scalebox{0.55}{2}}$}}
\psfrag{11}[][]{\scalebox{0.8}{$\,\,$: invariance axis}}
\psfrag{12}[][]{\scalebox{0.8}{$(\varepsilon_{\scalebox{0.55}{0}},\,\mu_{\scalebox{0.55}{0}})$}}
\psfrag{13}[][]{\scalebox{0.8}{Emitter}}
\psfrag{14}[][]{\scalebox{0.8}{Receiver}}
\psfrag{15}[][]{\scalebox{0.9}{$(\mathcal{V})$}}
\psfrag{16}[][]{\scalebox{0.9}{$(\Gamma)$}}
\includegraphics[height=8.0cm,width=11cm]{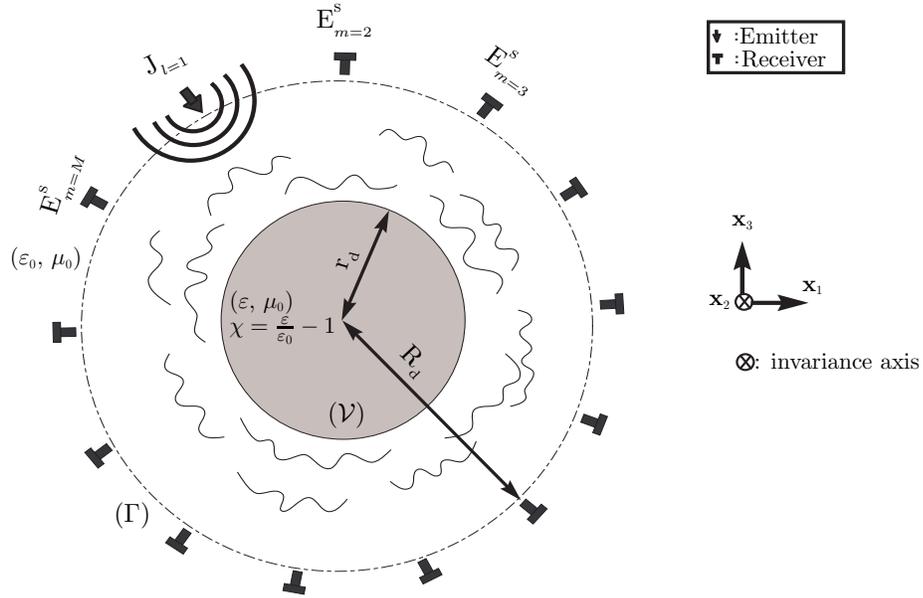} 
\caption{Scheme of the study configuration
\label{numerical-configuration}}
\end{psfrags}
\end{figure}

In Cartesian coordinates $({\rm O},\,{\mb x}_{\scalebox{0.55}{1}},\,{\mb x}_{\scalebox{0.55}{2}},\,{\mb x}_{\scalebox{0.55}{3}})$, we consider a cylindrical object of circular cross-section, and where ${\rm O}{\mb x}_{\scalebox{0.55}{2}}$ is the invariance axis (see figure \ref{numerical-configuration}). It is described by the complex permittivity-contrast function $\chi=\chi({\mb r})$, where ${\mb r}=x_{\scalebox{0.55}{1}}\,{\mb x}_{\scalebox{0.55}{1}}+x_{\scalebox{0.55}{3}}\,{\mb x}_{\scalebox{0.55}{3}}$ is a position vector in the plane perpendicular to the invariance axis. We work in ${\rm TE}$ polarization, and we model the electric component of the field parallel to the invariance axis. The total field calculated in the object, and the scattered field calculated at the receivers, are written respectively: ${\mb E}={\rm E}({\mb r})\,{\mb x}_{\scalebox{0.55}{2}}$ and ${\mb E}^{^{\scalebox{0.7}{\rm s}}}={\rm E}^{^{\scalebox{0.7}{\rm s}}}({\mb r}_{_m})\,{\mb x}_{\scalebox{0.55}{2}}$. We assume a harmonic time dependence of the form ${\rm e}^{-{\rm i}\omega t}$, with $\omega$ designating the frequency. In this study, we consider $M$ antennas aligned along a circle $(\Gamma)$ of radius ${\rm R_{_d}}$, and concentric with the object that has a radius ${\rm r_{_d}}$. Each antenna can act as an emitter or a receiver but not simultaneously. When an antenna $l$ acts as a source-current emitter oriented along the invariance axis ${\mb J}_{_l}={\rm J}_{_l}\,{\mb x}_{\scalebox{0.55}{2}}\,\,(l\in\{1,\,2,...,\,M\})$, all other antennas act as receivers to detect the scattered field ${\mb E}^{^{\scalebox{0.55}{\rm s}}}_{_m}$, $m\in\{1,\,2,...,\,M\}_{m\neq l}$. All antennas radiate with a source current of unit amplitude ${\rm J}_{_l}=1\,\,(l\in\{1,\,2,...,\,M\})$. An antenna located on the position ${\mb r}_{_l}=x^{_l}_{\scalebox{0.55}{1}}\,{\mb x}_{\scalebox{0.55}{1}}+x^{_l}_{\scalebox{0.55}{3}}\,{\mb x}_{\scalebox{0.55}{3}}$, radiates in the absence of the object the following incident field: \\

\begin{equation}\label{incidentfield}
{\rm E}^{^{\scalebox{0.7}{\rm i}}}_{_l}({\mb r})\,{\mb x}_{\scalebox{0.55}{2}}\,=\,-\frac{1}{4}\,\omega\,\mu_{\scalebox{0.55}{0}}\,\,{\rm H}^{^{{\scalebox{0.55}{(1)}}}}_{\scalebox{0.55}{0}}({\rm k}\,|{\mb r}-{\mb r}_{_l}|)\,\,{\mb x}_{\scalebox{0.55}{2}}
\vspace*{0.5cm}
\end{equation}

Here, ${\rm E}^{^{\scalebox{0.7}{\rm i}}}_{_l}({\mb r})$ is obtained by solving the scalar Helmholtz equation \cite{abramowitz1965handbook}, ${\rm k}=2\pi/\lambda$ is the wave vector module of the incident medium. ${\rm H}^{^{{\scalebox{0.55}{(1)}}}}_{\scalebox{0.55}{0}}$ is the zero-order first-kind Hankel function.

\subsection{Volume Integral Formulation}
\label{subsec:Volume Integral Formulation}
All vector physical quantities are aligned along the invariance axis. Therefore, the forward scattering problem addressed here is scalar and can be solved in the plane of incidence. This is done thanks to the volume integral formulation, which allows to build the forward model. The latter consists of the following two integral equations:

\begin{itemize}
\item[$\bullet$] \textbf{Sate equation} The expression (\ref{incidentfield}) allows to calculate the incident field ${\rm E}^{^{\scalebox{0.7}{\rm i}}}_{_l}$ everywhere inside the object. The total field ${\rm E}_{_l}$ inside the object is obtained by solving the following state equation:
\begin{equation}\label{statequation}
\begin{array}{l}
\forall{\mb r}\in\mathcal{V},\,\,{\rm E}_{_l}({\mb r})\,=\,{\rm E}^{^{\scalebox{0.7}{\rm i}}}_{_l}({\mb r})\,+\,({\rm i}\,{\rm k}^{\scalebox{0.55}{2}}/4)\,\int\limits_{\mathcal{V}}{\rm H}^{^{\scalebox{0.55}{(1)}}}_{\scalebox{0.55}{0}}({\rm k}\,|{\mb r}-{\mb r}{'}|)\,\chi({\mb r}{'})\,{\rm E}_{_l}({\mb r}{'})\,\,{\rm d}\mathcal{V}{'}
\end{array}
\end{equation}
\item[$\bullet$] \textbf{Observation equation} Once the total field ${\rm E}_{_l}$ is calculated using the state equation, it is replaced in the observation equation to model the scattered field ${\rm E}^{^{\scalebox{0.7}{\rm s}}}_{_l}$ on the receiver located on the position ${\mb r}_{_m}$:
\begin{equation}\label{observationequation}
\begin{array}{l}
\forall{\mb r}_{_m}\in\Gamma\,\,({\mb r}_{_m}\neq {\mb r}_{_l}),\,\,{\rm E}^{^{\scalebox{0.7}{\rm s}}}_{_l}({\mb r}_{_m})\,=\,({\rm i}\,{\rm k}^{\scalebox{0.55}{2}}/4)\,\int\limits_{\mathcal{V}}{\rm H}^{^{\scalebox{0.55}{(1)}}}_{\scalebox{0.55}{0}}({\rm k}\,|{\mb r}_{_m}-{\mb r}|)\,\chi({\mb r})\,{\rm E}_{_l}({\mb r})\,\,{\rm d}\mathcal{V}
\end{array}
\end{equation}
\item[$\bullet$] \textbf{Data of the scattered field} By varying the position ${\mb r}_{_l}$ of the source-current emitter ${\mb J}_{_l},\,\,l\in\{1,\,2,...,M\}$, and the position ${\mb r}_{_m}$, $m\in\{1,\,2,...,M\}_{m\neq l}$ of the receiver, we generate the data of the scattered field, noted: ${\rm E}^{^{\scalebox{0.7}{\rm s}}}\,=\,\{{\rm E}^{^{\scalebox{0.7}{\rm s}}}_{_{lm}}\,=\,{\rm E}^{^{\scalebox{0.7}{\rm s}}}_{_l}({\mb r}_{_m})\text{, for: }l,m\in\{1,\,2,...,\,M\}_{l\neq m}\}$. Equations (\ref{statequation},\ref{observationequation}) express a nonlinear link between the datum of the scattered field ${\rm E}^{^{\scalebox{0.7}{\rm s}}}_{_{lm}}(\chi)$ and the complex permittivity-contrast $\chi$. They are solved numerically using the moment method, the details of this numerical resolution are given in \cite{lesselier1991buried,tsang2004scatteringnum}.  
\end{itemize} 

\section{Inverse Problem Statement}
\label{sec:Inverse Problem Statement}

We have $M$ antennas, each one of them can operate in both emitter and receiver modes but not simultaneously, therefore, the data have their index $lm$ which takes the values $lm\,\in\{\,1,...,M^{\scalebox{0.55}{2}}-M\}$. In addition, the reciprocity theorem tells us that a $lm$ datum of the scattered field remains unchanged if we switch the position of the emitter with that of the receiver, which therefore reduces the number of distinct data to $(M^{\scalebox{0.55}{2}}-M)/2$. We note by ${\rm E}^{^{\scalebox{0.7}{\rm s,ex}}}\,=\,\{{\rm E}^{^{\scalebox{0.7}{\rm s,ex}}}_{_{lm}}\text{, for: }lm\in\{1,\,...,\,M^{\scalebox{0.55}{2}}-M\}\}$ the reference data of the actual object of complex permittivity-contrast $\chi^{\scalebox{0.55}{\rm ex}}$ that we are looking to reconstruct, and by ${\rm E}^{^{\scalebox{0.7}{\rm s}}}\,=\,\{{\rm E}^{^{\scalebox{0.7}{\rm s}}}_{_{lm}}\text{, for: }lm\in\{1,\,...,\,M^{\scalebox{0.55}{2}}-M\}\}$ the simulated data for an estimation $\chi$ of the complex permittivity-contrast of the object. Now, we can introduce the cost functional $\mathcal{F_{_C}}=\mathcal{F_{_C}}(\chi)\geq 0$, which expresses the normalized squared $\mathcal{L}_{\scalebox{0.55}{2}}$ norm of the discrepancy between reference data and simulated data: \\

\begin{equation}\label{costfunction}
\mathcal{F_{_C}}(\chi)\,=\,\frac{1}{\sum\limits_{lm=1}^{M^{\scalebox{0.55}{2}}-M}|{\rm E}^{^{\scalebox{0.7}{\rm s,ex}}}_{_{lm}}|^{^{\scalebox{0.55}{2}}}}\,\times\,\sum\limits_{lm=1}^{M^{\scalebox{0.55}{2}}-M}|{\rm E}^{^{\scalebox{0.7}{\rm s,ex}}}_{lm}\,-\,{\rm E}^{^{\scalebox{0.7}{\rm s}}}_{_{lm}}(\chi)|^{^{\scalebox{0.55}{2}}}
\end{equation}

The inverse problem is solved by updating the complex permittivity-contrast $\chi$, so that the simulated data ${\rm E}^{^{\scalebox{0.7}{\rm s}}}$ approaches the reference data ${\rm E}^{^{\scalebox{0.7}{\rm s,ex}}}$, in the sense of the cost function (\ref{costfunction}). \\

\section{Iterative Resolution}
\label{sec:Iterative Resolution}
To minimize (\ref{costfunction}), we build an iterative sequence $\chi_{_{n+1}}\,=\,\chi_{_n}\,+\,\delta\chi_{_n}$, with $\delta\chi_{_n}$ a descente direction calculated so as to ensure the decrease of the cost function $\mathcal{F_{_C}}(\chi_{_n})\,>\,\mathcal{F_{_C}}(\chi_{_{n+1}})$.

\subsection{Gauss-Newton Quadratic Model}
\label{subsec:Gauss-Newton Quadratic Model}
In the vicinity of the complex permittivity-contrast $\chi_{_n}$, the cost functional $\mathcal{F_{_C}}(\,\chi_{_n}\,+\,\zeta\,)$ can be locally approached by a quadratic model \cite{bertsekas1995} of the form : \\

\begin{equation}\label{modelquadap}
\left\lbrace\begin{array}{l}
\mathcal{F_{_C}}(\,\chi_{_n}\,+\,\zeta\,)\,\simeq\,\mathcal{F_{_C}}(\,\chi_{_n}\,)\,+\,\mathcal{M}^{^{\scalebox{0.55}{\rm GN}}}_{_n}(\,\zeta\,) \\
\\
\text{Where,}\\
\\
\mathcal{M}^{^{\scalebox{0.55}{\rm GN}}}_{_n}(\,\zeta\,)\,=\,\Re\{\zeta^{^{\scalebox{0.55}{\dag}}}\,\mathcal{G}_{_n}\}\,+\,\frac{1}{2}\,\zeta^{^{\scalebox{0.55}{\dag}}}\,\mathcal{H}^{^{\scalebox{0.55}{\rm GN}}}_{_n}\,\zeta,
\end{array}\right.
\end{equation}

with $\Re\{.\}$ which denotes the real part and $\dag$ a symbol that expresses the conjugated transpose, $\mathcal{G}_{_n}$ and $\mathcal{H}^{^{\scalebox{0.55}{\rm GN}}}_{_n}$ are respectively the complex gradient and the hermitian Hessian of Gauss-Newton, both calculated for the contrast $\chi_{_n}$. The explicit calculus of the model $\mathcal{M}^{^{\scalebox{0.55}{\rm GN}}}_{_n}(\,\zeta\,)$ is given in appendix \ref{gradhess}.

\subsection{Regularized Gauss-Newton Direction}
\label{subsec:Regularized Gauss-Newton Direction}
In the vicinity of the solution $\zeta_{\scalebox{0.55}{\rm op}}$ giving the minimum of the model $\mathcal{M}^{^{\scalebox{0.55}{\rm GN}}}_{_n}(\,\zeta\,)$, we have the equation : \\

\begin{equation}\label{variationmodel}
\left\lbrace\begin{array}{l}
\delta\mathcal{M}^{^{\scalebox{0.55}{\rm GN}}}_{_n}(\,\zeta_{\scalebox{0.55}{\rm op}}\,)\,=\,0\,\,\Leftrightarrow\,\,\Re\{\delta\zeta^{^{\scalebox{0.55}{\dag}}}_{\scalebox{0.55}{\rm op}}\,\mathcal{G}_{_n}\,+\,\delta\zeta^{^{\scalebox{0.55}{\dag}}}_{\scalebox{0.55}{\rm op}}\,\mathcal{H}^{^{\scalebox{0.55}{\rm GN}}}_{_n}\,\zeta_{\scalebox{0.55}{\rm op}}\}\,=\,0 \\
\\
\text{Which is generally satisfied when :}\\
\\
\mathcal{G}_{_n}\,+\,\mathcal{H}^{^{\scalebox{0.55}{\rm GN}}}_{_n}\,\zeta_{\scalebox{0.55}{\rm op}}\,=\,0
\end{array}\right.
\end{equation}

Let us call this solution Gauss-Newton direction and note it by   $\zeta_{\scalebox{0.55}{\rm op}}=\delta\chi^{\scalebox{0.55}{\rm GN}}_{_n}$. Its expression is written in the form: \\

\begin{equation}\label{modelquadapsolution}
\delta\chi^{\scalebox{0.55}{\rm GN}}_{_n}\,=\,-[\mathcal{H}^{^{\scalebox{0.55}{\rm GN}}}_{_n}]^{^{\scalebox{0.7}{-1}}}\,\mathcal{G}_{_n}
\end{equation}

The Hessian $\mathcal{H}^{^{\scalebox{0.55}{\rm GN}}}_{_n}$ is not defined positive, so it cannot be inverted. One way to deal with this difficulty is to look for the regularized Gauss-Newton direction of the following regularized Gauss-Newton quadratic model \cite{tikhonov1977solutions}: \\

\begin{equation}\label{modelquadapmod}
\mathcal{M}^{^{\widetilde{\scalebox{0.55}{\rm GN}}}}_{_n}(\,\zeta\,)\,=\,\Re\{\zeta^{^{\scalebox{0.55}{\dag}}}\,\mathcal{G}_{_n}\}\,+\,\frac{1}{2}\,\zeta^{^{\scalebox{0.55}{\dag}}}\,\mathcal{H}^{^{\scalebox{0.55}{\rm GN}}}_{_n}\,\zeta\,+\,\frac{\gamma_{_n}}{2}\,\zeta^{^{\scalebox{0.55}{\dag}}}\,\mathcal{I}\,\zeta,
\end{equation}

where $\mathcal{I}$ is the identity operator and $\gamma_{_n}$ the Tikhonov parameter which is strictly positive and which varies during iterations. In this case, the regularized Gauss-Newton direction, in the sense of the standard Tikhonov regularization is written: \\

\begin{equation}\label{modelquadapsolutionreg}
\delta\chi^{\widetilde{\scalebox{0.55}{\rm GN}}}_{_n}\,=\,-[\,\mathcal{H}^{^{\scalebox{0.55}{\rm GN}}}_{_n}\,+\,\gamma_{_n}\,\mathcal{I}\,]^{^{\scalebox{0.7}{-1}}}\,\mathcal{G}_{_n}
\end{equation}

The key point of this approach is to gradually decrease the value of the Tikhonov parameter over iterations. This gradually increases the sensitivity of the inverse operator  $[\,\mathcal{H}^{^{\scalebox{0.55}{\rm GN}}}_{_n}\,+\,\gamma_{_n}\,\mathcal{I}\,]^{^{\scalebox{0.7}{-1}}}$, with the result that the solution is reconstructed with more resolution. However, caution should be taken because a too low value of this regularization parameter may make the process unstable and lead to completely erroneous solutions, this is especially true for noisy data. Details of this iterative regularization are reported in the section dedicated to numerical results.

\subsection{Conjugate-Gradient Direction}
\label{subsec:Conjugate-Gradient Direction}
The Conjugate-Gradient direction is an approximate solution to the problem of minimizing the Gauss-Newton quadratic model $\mathcal{M}^{^{\scalebox{0.55}{\rm GN}}}_{_n}$. This descent direction that we note $\delta\chi^{\scalebox{0.55}{\rm CG}}_{_n}$ is written: \\

\begin{equation}\label{solutioncg}
\delta\chi^{\scalebox{0.55}{\rm CG}}_{_n}\,=\,\alpha_{_n}\,\mathcal{D}_{_n}
\end{equation}

With $\mathcal{D}_{_n}$ which is given by the Polak-Ribi\`ere formulation: \\

\begin{equation}\label{formulepolakrib}
\left\lbrace\begin{array}{l}
\mathcal{D}_{_n}\,=\,-\mathcal{G}_{_n},\,\,\,\text{for }n=0\\
\\
\mathcal{D}_{_n}\,=\,-\mathcal{G}_{_n}\,+\,\frac{\mathcal{G}^{^{\scalebox{0.55}{\dag}}}_{_n}\,\mathcal{H}^{^{\scalebox{0.55}{\rm GN}}}_{_n}\,\mathcal{D}_{_{n-1}}}{\mathcal{D}^{^{\scalebox{0.55}{\dag}}}_{_{n-1}}\,\mathcal{H}^{^{\scalebox{0.55}{\rm GN}}}_{_n}\,\mathcal{D}_{_{n-1}}}\,\,\mathcal{D}_{_{n-1}},\,\,\,\text{for }n \geq 1\\
\end{array}\right.
\end{equation}

The scale parameter $\alpha_{_n}$ is real, it is obtained by approaching the solution minimizing the quadratic model (\ref{modelquadap}) by the Conjugate-Gradient direction (\ref{solutioncg}) \cite{bertsekas1995}. This leads to the following equation, which is sufficient to determine this parameter: \\

\begin{equation}
\frac{{\rm d}\mathcal{M}^{^{\scalebox{0.55}{\rm GN}}}_{_n}(\,\alpha_{_n}\,\mathcal{D}_{_n}\,)}{{\rm d}\alpha_{_n}}\,=\,0
\end{equation}

Whose solution is written: \\

\begin{equation}\label{paramdechelle}
\alpha_{_n}\,=\,-\Re\{\frac{\mathcal{D}^{^{\scalebox{0.55}{\dag}}}_{_n}\,\mathcal{G}_{_n}}{\mathcal{D}^{^{\scalebox{0.55}{\dag}}}_{_n}\,\mathcal{H}^{^{\scalebox{0.55}{\rm GN}}}_{_n}\,\mathcal{D}_{_n}}\},
\end{equation}

with $\Re\{\}$ which refers to the real part of a complex number. The expressions (\ref{formulepolakrib}, \ref{paramdechelle}) summarize the Conjugate-Gradient iterative scheme. As it is based on an approximate solution to the problem of minimizing the Gauss-Newton quadratic model, it is characterized by a lower resolution and convergence rate than the iterative Gauss-Newton scheme. However, since it does not require inverting an operator, it remains faster in terms of computing time per iteration. In addition, it is inherently robust to noise, and completely autonomous since the scale parameter is calculated by the above analytical formula.

\section{Numerical Study}
\label{sec:Numerical Study}

\begin{figure}[h]
\centering
\begin{psfrags}
\psfrag{t1}[][]{{\color{white}\scalebox{0.75}{$\varepsilon_{_{\rm r}}=4,\,\sigma=0\,\,{\rm S.m^{\scalebox{0.55}{-1}}}$}}}
\psfrag{t2}[][]{{\color{white}\scalebox{0.75}{$\varepsilon_{_{\rm r}}=4,\,\sigma=0.1\,\,{\rm S.m^{\scalebox{0.55}{-1}}}$}}}
\psfrag{t3}[][]{{\color{white}\scalebox{0.75}{$\varepsilon_{_{\rm r}}=5,\,\sigma=0.1\,\,{\rm S.m^{\scalebox{0.55}{-1}}}$}}}
\psfrag{t4}[][]{{\color{white}\scalebox{0.75}{$\varepsilon_{_{\rm r}}=5,\,\sigma=0\,\,{\rm S.m^{\scalebox{0.55}{-1}}}$}}}
\includegraphics[height=8.5cm,width=10cm]{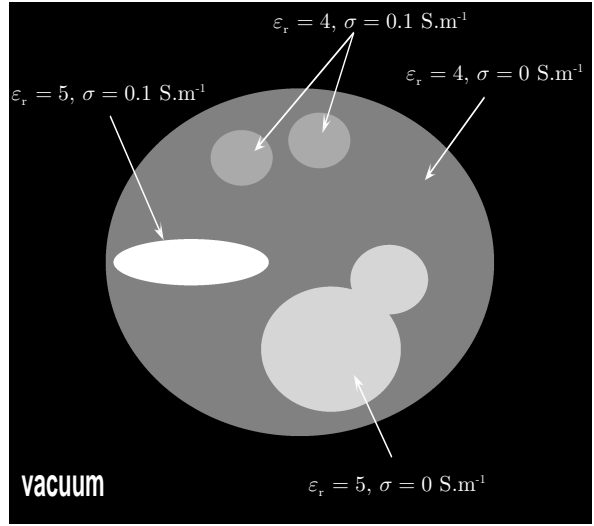} \\
\caption{The actual object we are reconstructing, its initial guess introduced into the inversion process, is reduced to considering a zero conductivity $\sigma=0\,\,{\rm S.m^{\protect\scalebox{0.55}{-1}}}$ and a relative dielectirc permittivity equal to $\varepsilon_{_{\rm r}}=4$, everywhere in the domain $\mathcal{V}$. 
\label{fig1}}
\end{psfrags}
\end{figure}

\begin{figure}[h]
\centering
\includegraphics[height=6.5cm,width=10cm]{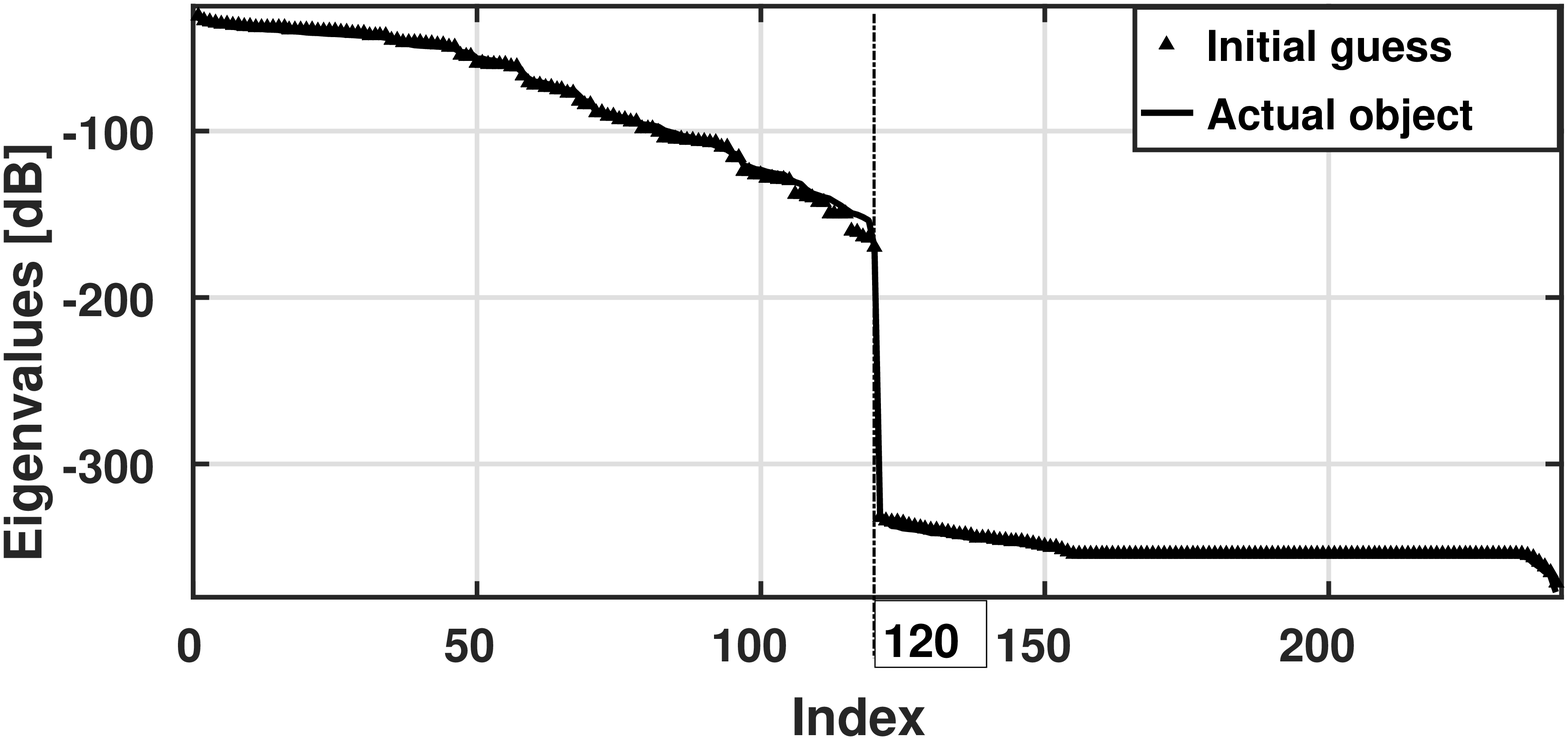} \\
\caption{Distribution in decibels $[{\rm dB}]$ of the non-zero eigenvalues of operator $\mathcal{H}^{^{\protect\scalebox{0.55}{\rm GN}}}_{_{\rm ex}}$ of the actual object, and of the eigenvalues of operator $\mathcal{H}^{^{\protect\scalebox{0.55}{\rm GN}}}_{\protect\scalebox{0.55}{0}}$ of its initial guess. 
\label{fig2}}
\end{figure}

\subsection{Numerical Parameters}
\label{subsec:Numerical Parameters}
To simulate the inversion process, the domain $\mathcal{V}$ of the object, of circular shape, of radius ${\rm r}_{_{\rm d}}=5\,{\rm cm}$ is discretized with ${\rm Q}=7764$ pixels, each pixel has a side $\Delta\simeq{1.01\,{\rm mm}}$ and an area $\Delta\mathcal{V}\simeq 1.01\,\,{\rm mm}^{\scalebox{0.55}{2}}$. As an initial guess of the inversion process, we choose the object of the figure \ref{fig1} without its inclusions so with a relative dielectric permittivity of $\varepsilon_{_{\rm r}}=4$ and a zero conductivity $\sigma=0\,\,{\rm S.m^{\scalebox{0.55}{-1}}}$, everywhere in the domain $\mathcal{V}$. To avoid an inverse crime, the reference data that are inverted are generated by discretizing the object with a number of pixels equal to ${\rm Q}_{_{\rm r}}=9968$, corresponding to a side $\Delta_{_{\rm r}}\simeq 0.89\,\,{\rm mm}$ and an area $\Delta\mathcal{V}_{_{\rm r}}\simeq 0.79\,{\rm mm}^{\scalebox{0.55}{2}}$. The object that we invert and for which we generate the reference data is illustrated in figure \ref{fig1}, it has four inclusions, two of these inclusions are circular and very close to each other, they have a relative dielectric permittivity identical to that of the initial guess of the object $\varepsilon_{_{\rm r}}=4$, and have a conductivity $\sigma=0.1\,\,{\rm S.m^{\scalebox{0.55}{-1}}}$. Then, there is an oval inclusion which is of relative dielectric permittivity $\varepsilon_{_{\rm r}}=5$ and conductivity $\sigma=0.1\,{\rm S.m^{\scalebox{0.55}{-1}}}$. The fourth inclusion that looks like a pear has a zero conductivity $\sigma=0\,\,{\rm S.m^{\scalebox{0.55}{-1}}}$ and a relative dielectric permittivity $\varepsilon_{_{\rm r}}=5$. The antennas with the number $M=16$ are arranged circularly at an equal distance ${\rm R}_{_{\rm d}}\,=\,40\,{\rm cm}$ from the center of the object, for a total number of data equal to $240$, where due to reciprocity each data is repeated twice, reducing the number of distinct data to $120$. The wavelength of the incident field is set at $\lambda=10\,\,{\rm cm}$, corresponding to a frequency of $3\,\,{\rm GHz}$. Note that $\varepsilon_{_{\rm r}}$ and $\sigma$ are related to the complex permittivity-contrast by the relations: \\

\begin{equation}\label{chiepsilonsigma}
\left\lbrace\begin{array}{l}
\varepsilon_{_{\rm r}}\,=\,\Re\{\chi\}\,+\,1 \\
\\
\sigma\,=\,\omega\,\varepsilon_{\scalebox{0.55}{0}}\,\Im\{\chi\},
\end{array}\right.
\end{equation}

where $\Re\{.\}$ and $\Im\{.\}$ refer respectively to the real and imaginary parts of a complex number.

\subsection{Iterative Regularization of Gauss-Newton Direction}
\label{subsec:Iterative Regularization of Gauss-Newton Direction}
This iterative regularization is based on a study of the distribution curve of the nonzero eigenvalues of the operator  $\mathcal{H}^{^{\scalebox{0.55}{\rm GN}}}_{\scalebox{0.55}{0}}$ calculated for the initial guess of the object, and the operator $\mathcal{H}^{^{\scalebox{0.55}{\rm GN}}}_{_{\rm ex}}$ calculated for the actual object. The figure \ref{fig2} illustrates in decibels the distribution of the eigenvalues of $\mathcal{H}^{^{\scalebox{0.55}{\rm GN}}}_{\scalebox{0.55}{0}}$ and $\mathcal{H}^{^{\scalebox{0.55}{\rm GN}}}_{_{\rm ex}}$, we notice that on this scale, the two distributions differ slightly from each other, this means that during a reconstruction that starts from the initial guess of the object and evolves to get as close as possible to the actual object, the distribution curve of the eigenvalues of the Hessian evolves slightly on the decibel scale. This remarkable observation allows us to use the eigenvalues of $\mathcal{H}^{^{\scalebox{0.55}{\rm GN}}}_{\scalebox{0.55}{0}}$ to update the $\gamma_{_n}$ parameter during the iterations. This numerical point of view remains similar to the Tikhonov regularization and the truncated singular values decomposition \cite{hansen1990truncated}, but with an adjustable regularization parameter. Let be $\xi_{_p}^{^n}$ et $\lambda_{_p}^{^n}$ the $p$ ith eigenvector and eigenvalue of the operator $\mathcal{H}^{^{\scalebox{0.55}{\rm GN}}}_{_n}$, then the solution $\delta\chi^{\widetilde{\scalebox{0.55}{\rm GN}}}_{_n}$ expressed in (\ref{modelquadapsolutionreg}) can be decomposed in an equivalent way in the form: \\

\begin{equation}\label{modelquadapsolutiondecomp}
\delta\chi^{\widetilde{\scalebox{0.55}{\rm GN}}}_{_n}\,=\,\sum\limits_{p=1}^{\rm Q}\frac{<\xi_{_p}^{^n}|\mathcal{G}_{_n}>}{\gamma_{_n}\,+\,\lambda_{_p}^{^n}}\,|\xi_{_p}^{^n}>,
\end{equation}

with $<.|.>$ which refers to the inner product. The eigenvalues of $\mathcal{H}^{^{\scalebox{0.55}{\rm GN}}}_{_n}$ are distributed in decreasing order as follows: $\lambda_{_1}^{^n} > \lambda_{_2}^{^n}>\lambda_{_3}^{^n}>...\geq 0$. The numerical trick of this iterative regularization is to assign to the parameter $\gamma_{_n}$, the value of the $n$ ith eigenvalue $\lambda_{_n}^{\scalebox{0.55}{0}}$ of the operator $\mathcal{H}^{\scalebox{0.55}{\rm GN}}_{_0}$, whose eigenvalues are always distributed in the same decreasing order $\lambda_{_1}^{\scalebox{0.55}{0}} > \lambda_{_2}^{\scalebox{0.55}{0}}>\lambda_{_3}^{\scalebox{0.55}{0}}>...\geq 0$. This allows the solution (\ref{modelquadapsolutiondecomp}) to be rewritten as: \\

\begin{equation}\label{modelquadapsolutiondecompp}
\delta\chi^{\widetilde{\scalebox{0.55}{\rm GN}}}_{_n}\,=\,\sum\limits_{p=1}^{\rm Q}\frac{<\xi_{_p}^{^n}|\mathcal{G}_{_n}>}{\lambda_{_n}^{\scalebox{0.55}{0}}\,+\,\lambda_{_p}^{^n}}\,|\xi_{_p}^{^n}>
\end{equation}

This technique gradually reduces the value of the Tikhonov parameter over iterations, drawing its values from the spectrum of the eigenvalues of the operator $\mathcal{H}^{^{\scalebox{0.55}{\rm GN}}}_{\scalebox{0.55}{0}}$. It has the advantage of gradually increasing the sensitivity of the inversion process to gain in precision, however, care must be taken to choose judiciously the maximum number $N$ of iterations to avoid too low values of the Tikhonov parameter, for which we have an unstable iterative scheme and erroneous reconstructed solutions. This regularization technique also offers the flexibility to parsimoniously exploit the eigenvalues of $\mathcal{H}^{^{\scalebox{0.55}{\rm GN}}}_{\scalebox{0.55}{0}}$, this is done by replacing in the expression (\ref{modelquadapsolutiondecompp}) $\lambda_{_n}^{\scalebox{0.55}{0}}$ by $\lambda_{_{1\,+\,\kappa(n-1)}}^{\scalebox{0.55}{0}}$, where $\kappa$ is the index shift between two successive eigenvalues. The numerical results of this work were obtained by setting this parameter to the value $\kappa=1$.

\subsection{Monitoring Criteria}
\label{subsec:Monitoring Criteria}
To monitor the inversion process, we introduce two criteria expressed on the decibel scale, the first calculates the ratio of the squared $\mathcal{L}_{\scalebox{0.55}{2}}$ norm of the discrepancy between the data simulated at each iteration and the reference data, and the squared $\mathcal{L}_{\scalebox{0.55}{2}}$ norm of the reference data, the second criterion calculates the ratio of the squared $\mathcal{L}_{\scalebox{0.55}{2}}$ norm of the difference between the reconstructed object at each iteration and the actual object, and the squared of the $\mathcal{L}_{\scalebox{0.55}{2}}$ norm of the actual object. These two criteria, which are referred to as the residual error on the field and the residual error on the contrast, are given by their respective expressions: \\

\begin{equation}\label{monitoringcriteria}
\left\lbrace\begin{array}{l}
{\rm R}_{_{\rm ef}}(n)\,=\,10\,\times\,\log_{_{10}}\,\left(\frac{1}{\sum\limits_{lm=1}^{M^{\scalebox{0.55}{2}}-M}|{\rm E}^{^{\scalebox{0.55}{\rm s,ex}}}_{_{lm}}|^{^{\scalebox{0.55}{2}}}}\,\times\sum\limits_{lm=1}^{M^{\scalebox{0.55}{2}}-M}|{\rm E}^{^{\scalebox{0.7}{\rm s,ex}}}_{_{lm}}-{\rm E}^{^{\scalebox{0.7}{\rm s}}}_{_{lm}}(\chi_{_n})|^{^{\scalebox{0.55}{2}}}\right) \\
\\
{\rm R}_{_{\rm ep}}(n)\,=\,10\,\times\,\log_{_{10}}\,\left(\frac{1}{\sum\limits_{q=1}^{Q}|\chi^{\scalebox{0.55}{\rm ex}}_{_q}|^{^{\scalebox{0.55}{2}}}}\,\times\sum\limits_{q=1}^{Q}|\chi^{\scalebox{0.55}{\rm ex}}_{_q}-\chi_{_q}(n)|^{^{\scalebox{0.55}{2}}}\right),
\end{array}\right.
\end{equation}

with $\chi_{_q}(n)$ the $q$ th element of $\chi_{_n}$ which is the estimation of the complex permittivity contrast at the $n$ th step of the iterative process.

\subsection{Inversion of Noiseless Data}
\label{subsec:Inversion of Noiseless Data}
\begin{figure}[h]
\centering
\begin{tabular}{cccc}
\includegraphics[height=5.5cm,width=8.0cm]{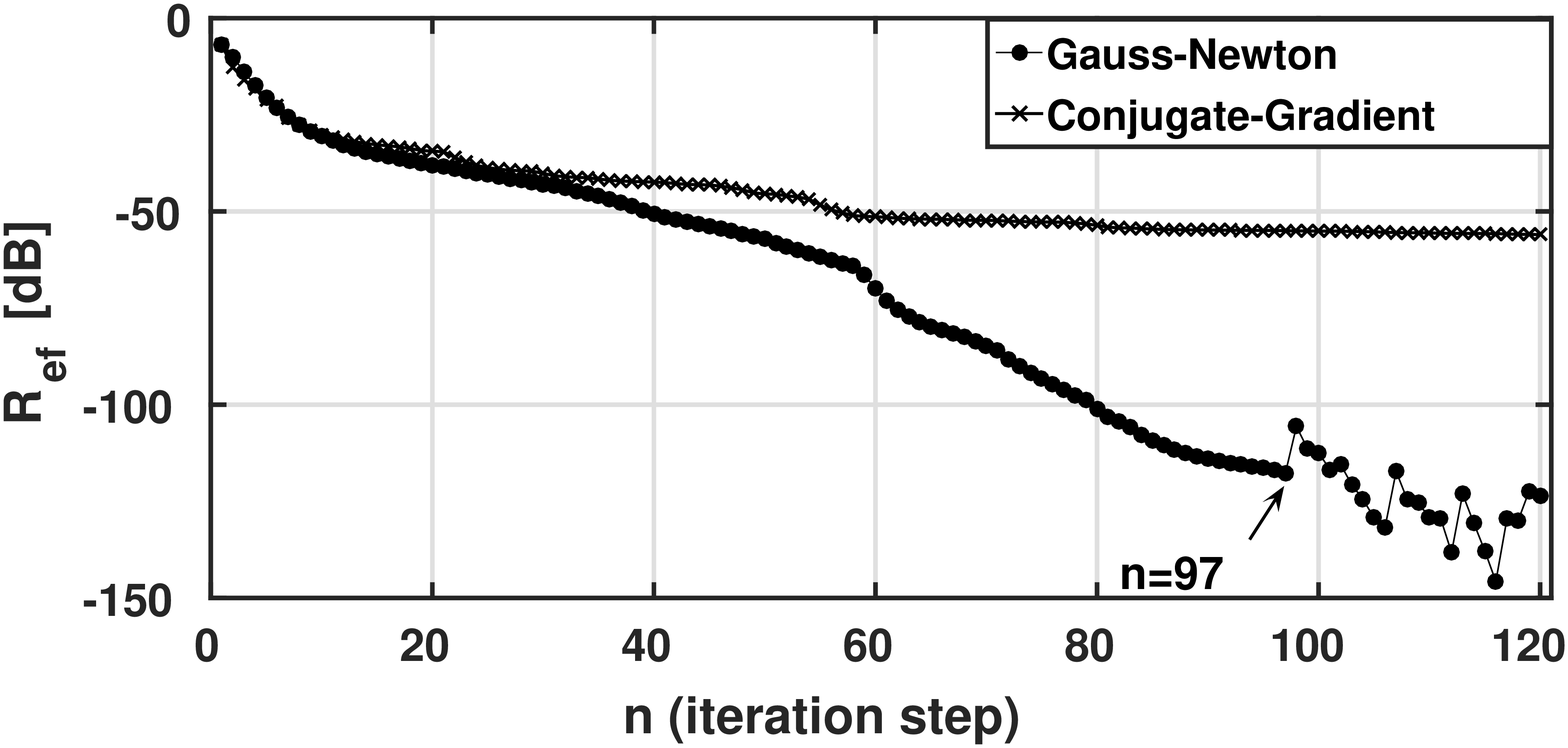} & & & \includegraphics[height=5.5cm,width=8.0cm]{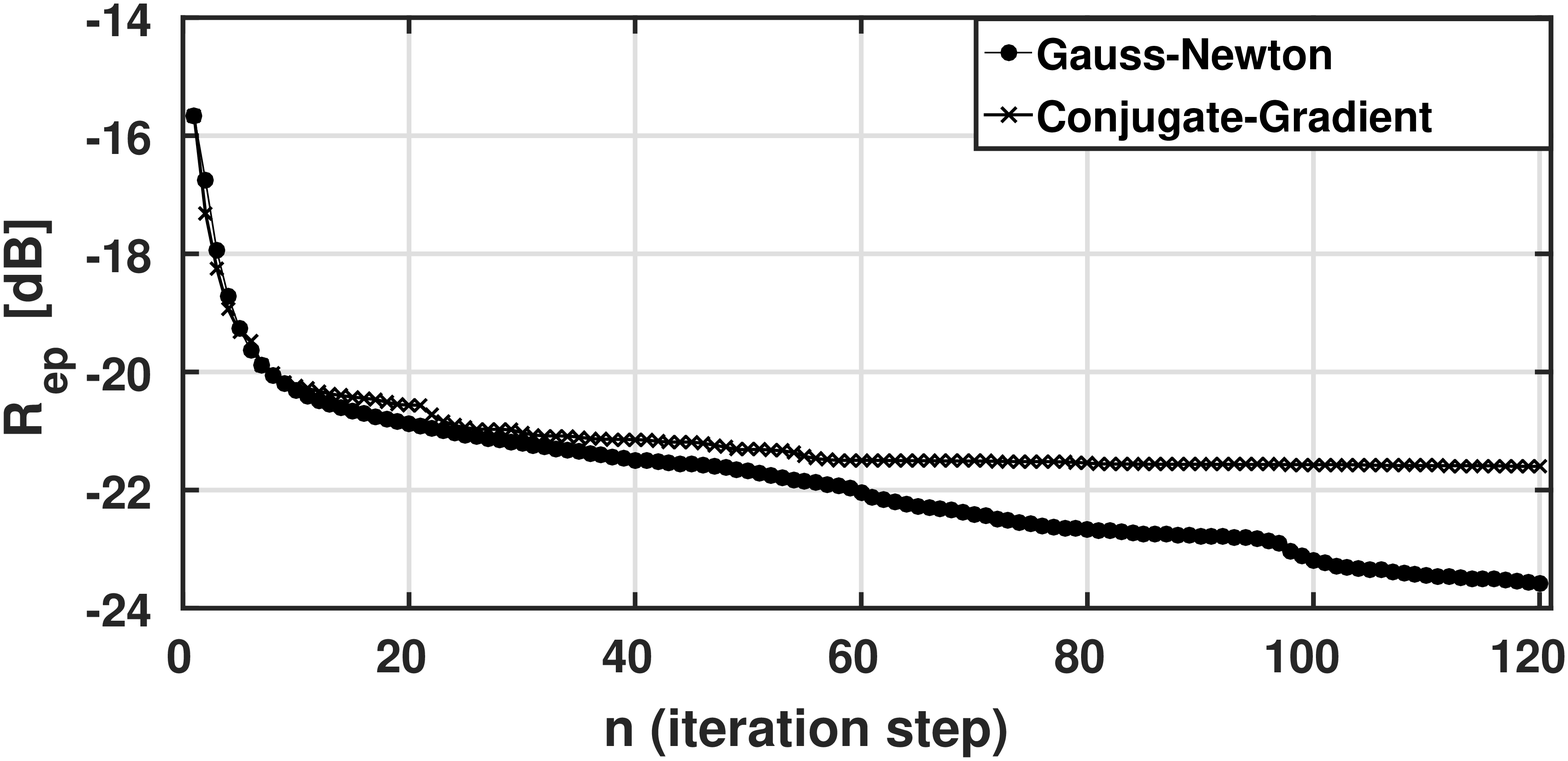} \\
(a) & & & (b) \\
\end{tabular}
\caption{(a): Behaviour of residual errors on the data, calculated in decibels $[{\rm dB}]$ for regularized Gauss-Newton and Conjugate-Gradient reconstructions, according to the number of iterations. (b): Behaviour of residual errors on the complex permittivity-contrast, calculated in decibels $[{\rm dB}]$ for regularized Gauss-Newton and Conjugate-Gradient reconstructions, according to the number of iterations.
\label{fig34}}
\end{figure}

\begin{figure}[h]
\centering
\begin{tabular}{cccc}
\includegraphics[height=6.5cm,width=7cm]{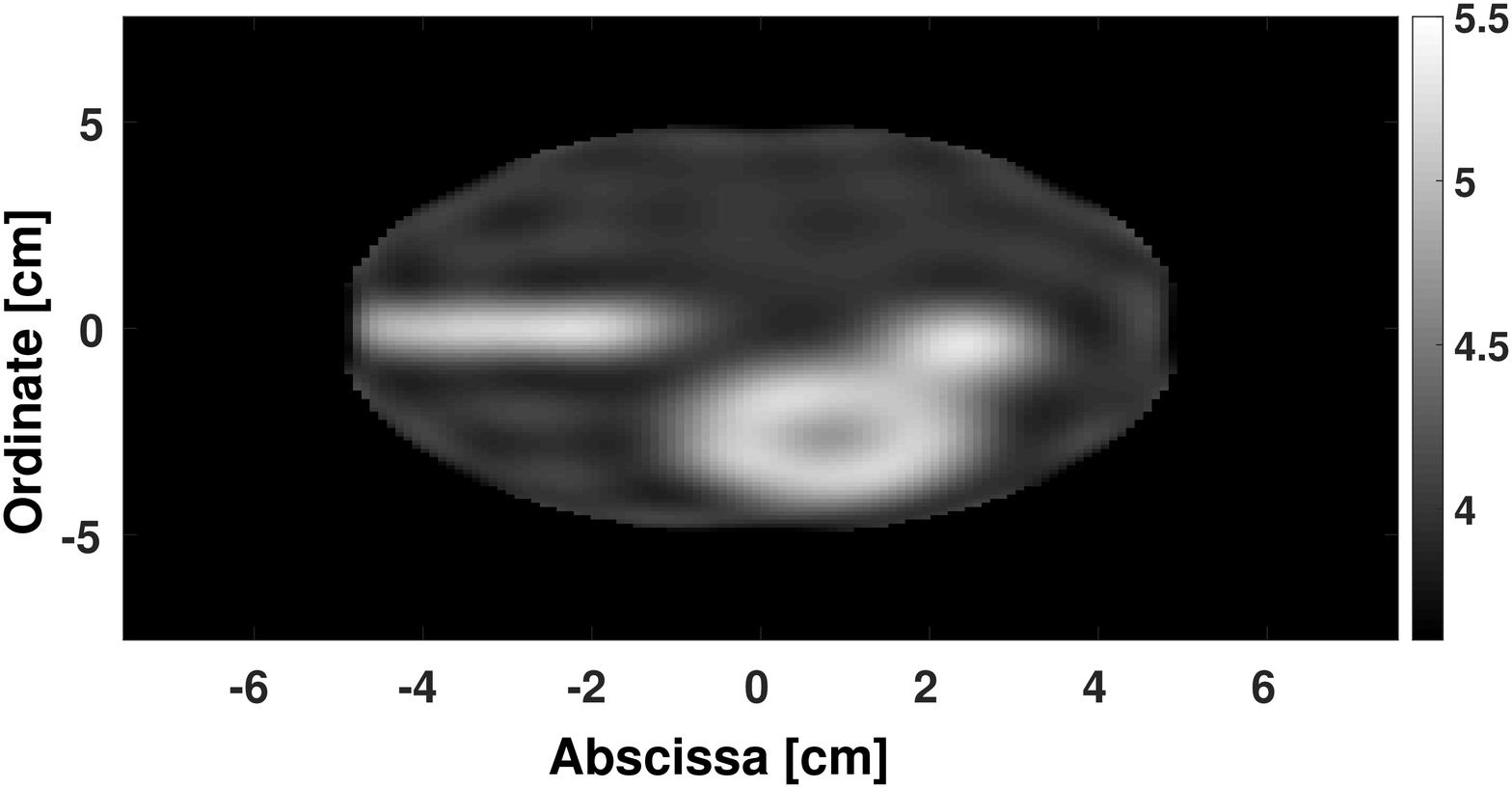} & & & \includegraphics[height=6.5cm,width=7cm]{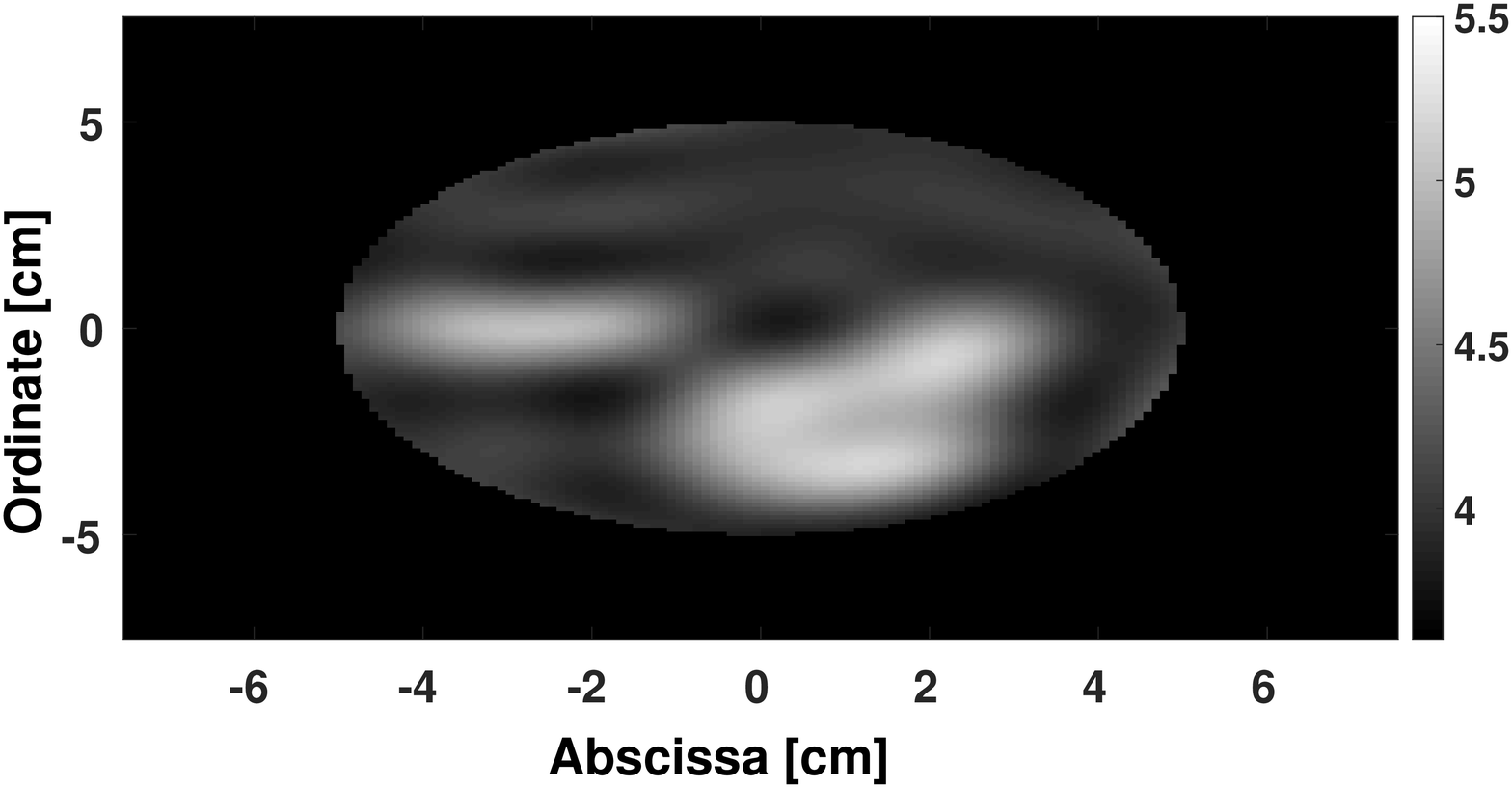} \\
(a) & & & (b) \\
\includegraphics[height=6.5cm,width=7cm]{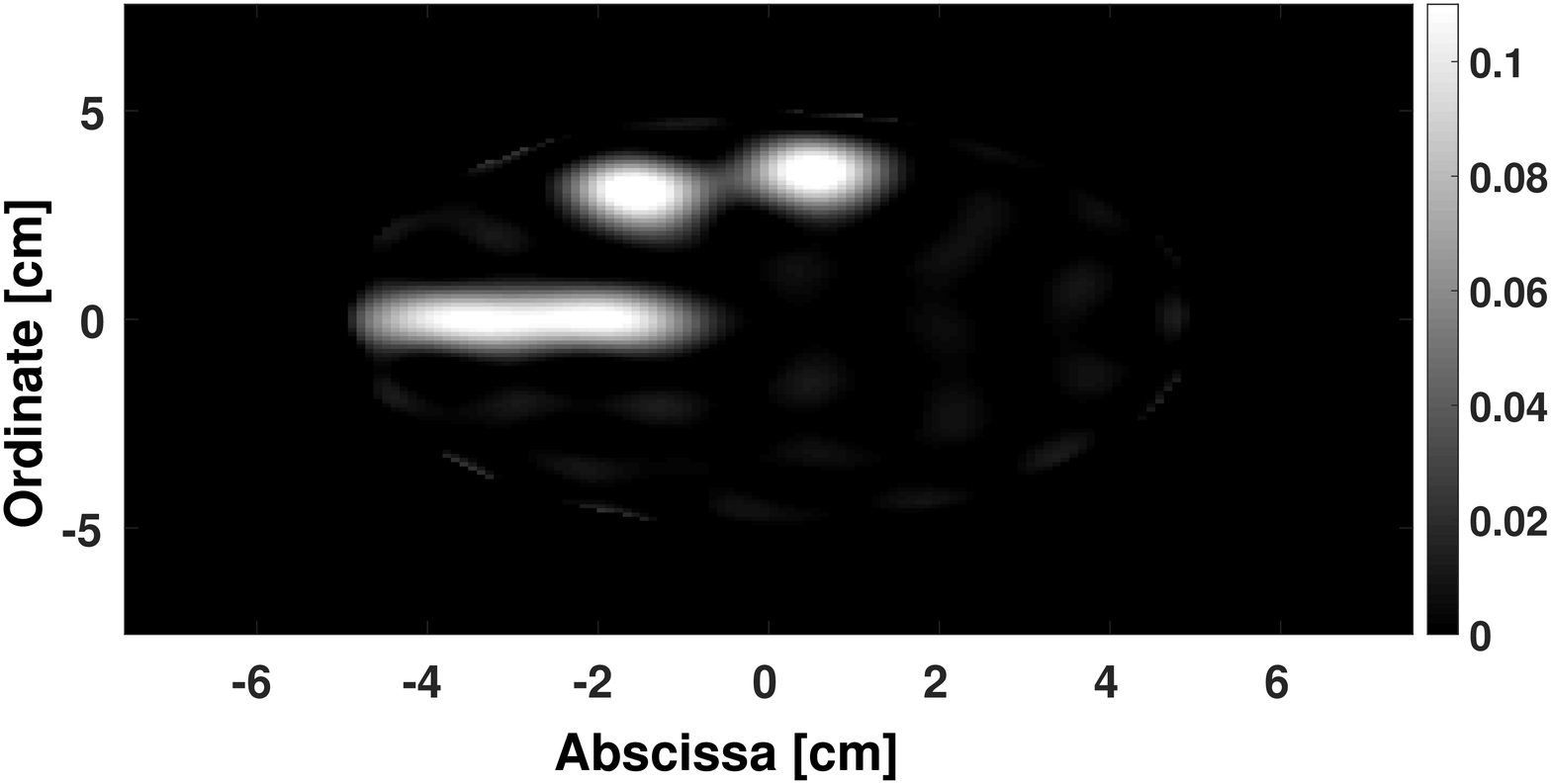} & & & \includegraphics[height=6.5cm,width=7cm]{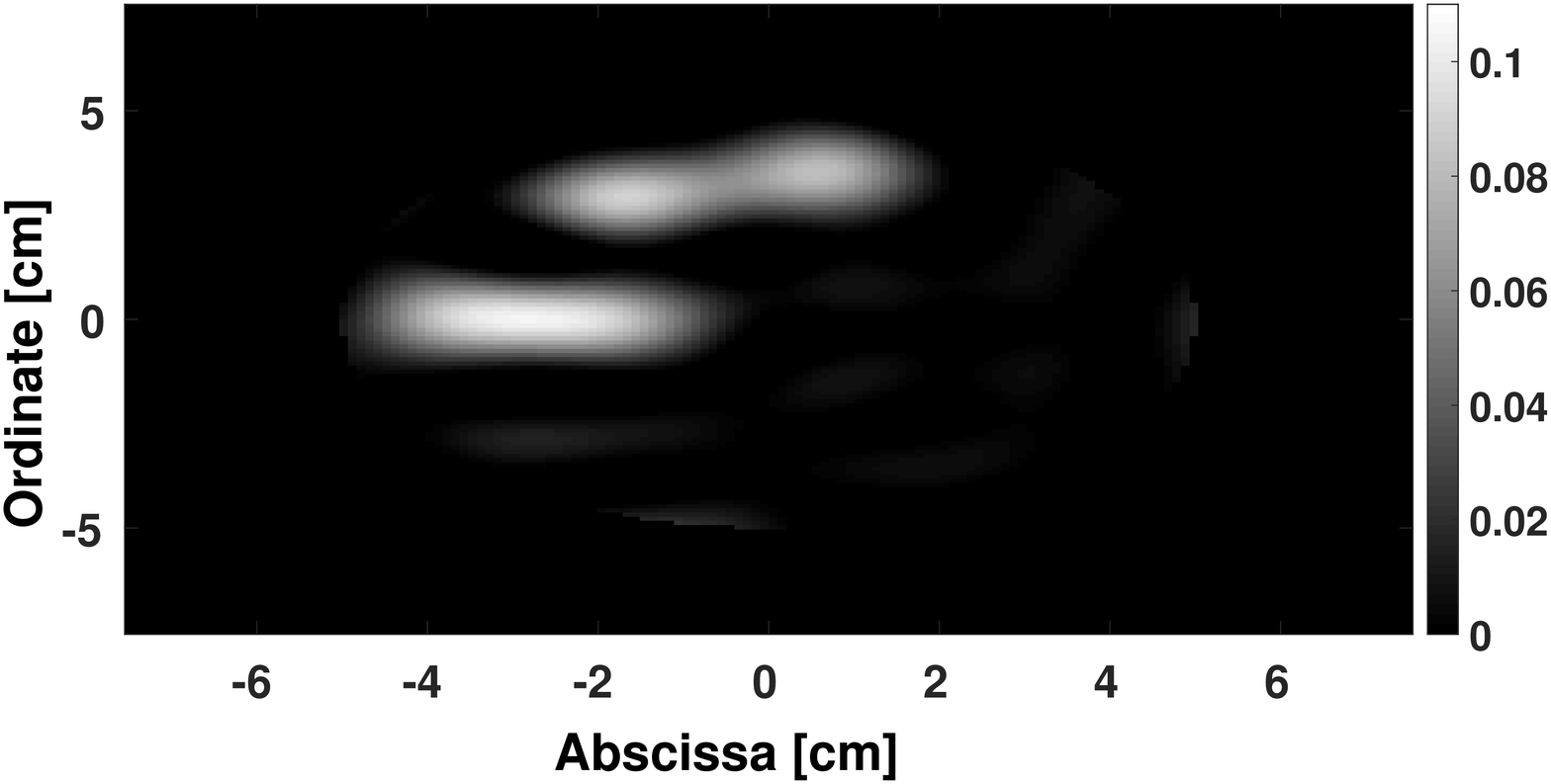} \\
(c) & & & (d) \\
\end{tabular}
\caption{(a): relative permittivity $\varepsilon_{_{{{\rm r}}}}^{\widetilde{\protect\scalebox{0.55}{\rm GN}}}(n=120)$ reconstructed after 120 iterations with the regularized Gauss-Newton. (b): relative permittivity $\varepsilon_{_{{{\rm r}}}}^{\protect\scalebox{0.55}{\rm CG}}(n=120)$ reconstructed after 120 iterations with the Conjugate-Gradient. (c): conductivity $\sigma^{\widetilde{\protect\scalebox{0.55}{\rm GN}}}(n=120)$ $({\rm S.m^{\protect\scalebox{0.55}{-1}}})$ reconstructed after 120 iterations with the regularized Gauss-Newton. (d): conductivity $\sigma^{\protect\scalebox{0.55}{\rm CG}}(n=120)$ $({\rm S.m^{\protect\scalebox{0.55}{-1}}})$ reconstructed after 120 iterations with the Conjugate-Gradient.
\label{fig5678}}
\end{figure}

\begin{figure}[h]
\centering
\begin{tabular}{cccc}
\includegraphics[height=6.5cm,width=7cm]{fig5.eps} & & & \includegraphics[height=6.5cm,width=7cm]{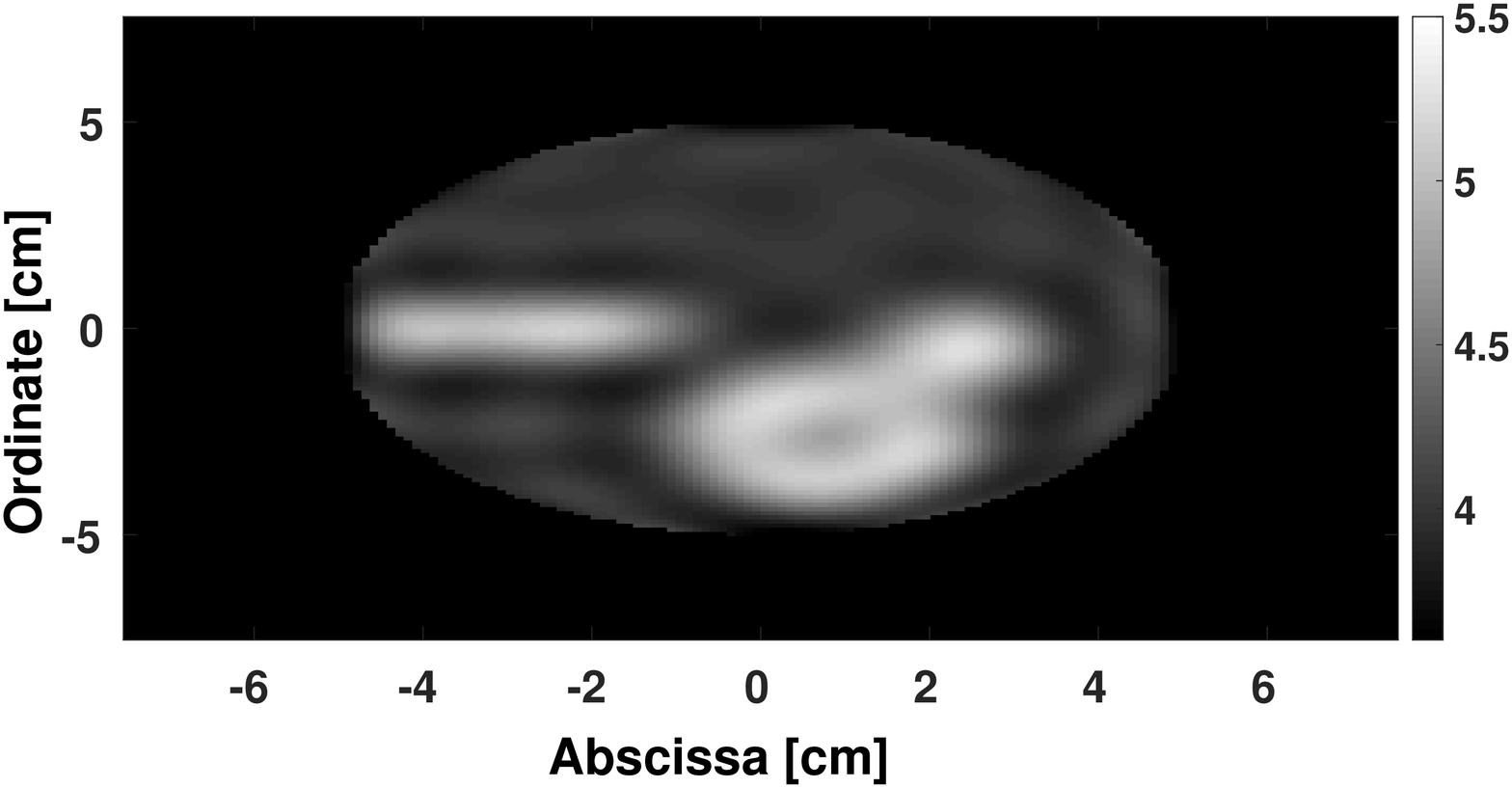} \\
(a) & & & (b) \\
\includegraphics[height=6.5cm,width=7cm]{fig7.eps} & & & \includegraphics[height=6.5cm,width=7cm]{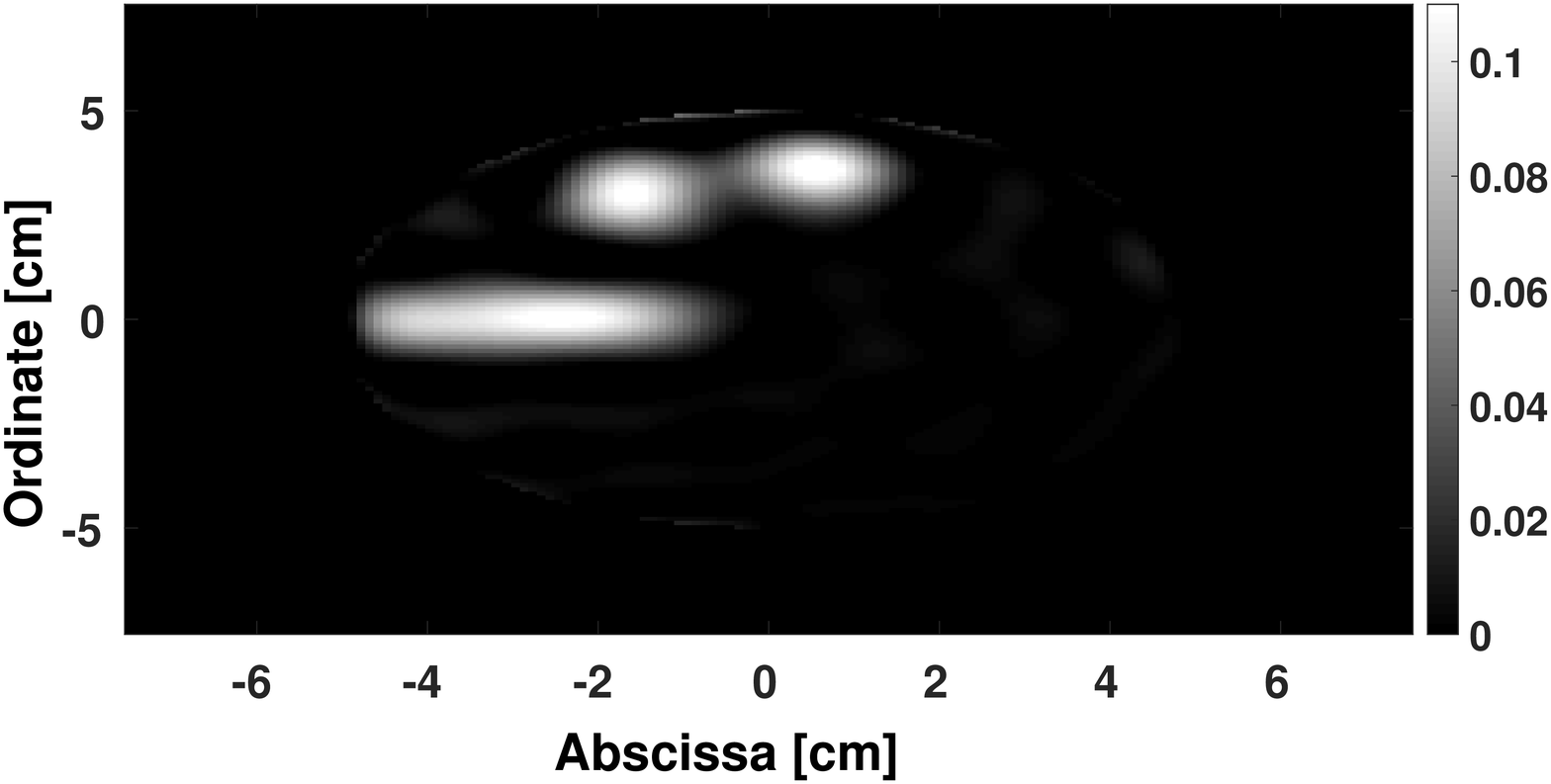} \\
(c) & & & (d) \\
\end{tabular}
\caption{(a): relative permittivity $\varepsilon_{_{{{\rm r}}}}^{\widetilde{\protect\scalebox{0.55}{\rm GN}}}(n=120)$ reconstructed after 120 iterations with the regularized Gauss-Newton. (b): relative permittivity $\varepsilon_{_{{{\rm r}}}}^{\widetilde{\protect\scalebox{0.55}{\rm GN}}}(n=97)$ reconstructed after 97 iterations with the regularized Gauss-Newton. (c): conductivity $\sigma^{\widetilde{\protect\scalebox{0.55}{\rm GN}}}(n=120)$ $({\rm S.m^{\protect\scalebox{0.55}{-1}}})$ reconstructed after 120 iterations with the regularized Gauss-Newton. (d): conductivity $\sigma^{\widetilde{\protect\scalebox{0.55}{\rm GN}}}(n=97)$ $({\rm S.m^{\protect\scalebox{0.55}{-1}}})$ reconstructed after 97 iterations with the regularized Gauss-Newton.
\label{fig59710}}
\end{figure}

Figure \ref{fig34}-(a) illustrates on the decibel scale, the behaviour of the residual error on the field calculated for the regularized Gauss-Newton and Conjugate-Gradient iterative schemes, as a function of the number of iterations. Figure \ref{fig34}-(b) uses the same scale, and compares these two approaches on the evolution of the residual error on complex permittivity-contrast, as a function of the number of iterations. For both approaches, the monitoring criteria $({\rm R}_{_{\rm ef}},\,{\rm R}_{_{\rm ep}})$ decrease over the iterations, with a higher convergence rate for the regularized Gauss-Newton iterative scheme than for the Conjugate-Gradient iterative scheme. Concerning the calculation time per iteration, the Conjugate-Gradient iterative scheme consumes $33.4\%$ less time than the regularized Gauss-Newton iterative scheme, this is due to the fact that the calculus at each iteration of the Conjugate-Gradient descent direction does not require a matrix inversion as it is the case with the regularized Gauss-Newton iterative scheme. The solutions obtained after $120$ iterations, show that the relative permittivity (see figures \ref{fig5678}-(a) and \ref{fig5678}-(b)) of the inclusions and their conductivity (see figs \ref{fig5678}-(c) and \ref{fig5678}-(d)) are reconstructed with better resolution by the regularized Gauss-Newton iteration scheme. In figure \ref{fig34}-(a) we note that starting from iteration 98, the emergence of oscillations in the curve describing the behaviour of the ${\rm R}_{_{\rm ef}}$ criterion for the regularized Gauss-Newton iterative scheme. This instability is due to the combined actions of noise from numerical rounding errors and a regularization parameter that becomes lower and lower during iterations. A comparison between the solution of the regularized Gauss-Newton iterative scheme extracted at iteration 97 and the one we obtain at iteration $120$, shows (see figures \ref{fig59710}-(a), \ref{fig59710}-(b), \ref{fig59710}-(c) and \ref{fig59710}-(d)) that they are very similar to each other. For indices above $120$, there is a brutal decrease in the eigenvalues of $\mathcal{H}^{^{\scalebox{0.55}{\rm GN}}}_{\scalebox{0.55}{0}}$ (see figure \ref{fig2}), therefore, attributing to the regularization parameter such very low eigenvalues from the range with indices above $120$, leads to an unstable regularized Gauss-Newton iterative scheme. 

\subsection{Inversion of Noisy Data}
\label{subsec:Inversion of Noisy Data}
\begin{figure}[h]
\centering
\begin{tabular}{cccc}
\includegraphics[height=5.5cm,width=8.0cm]{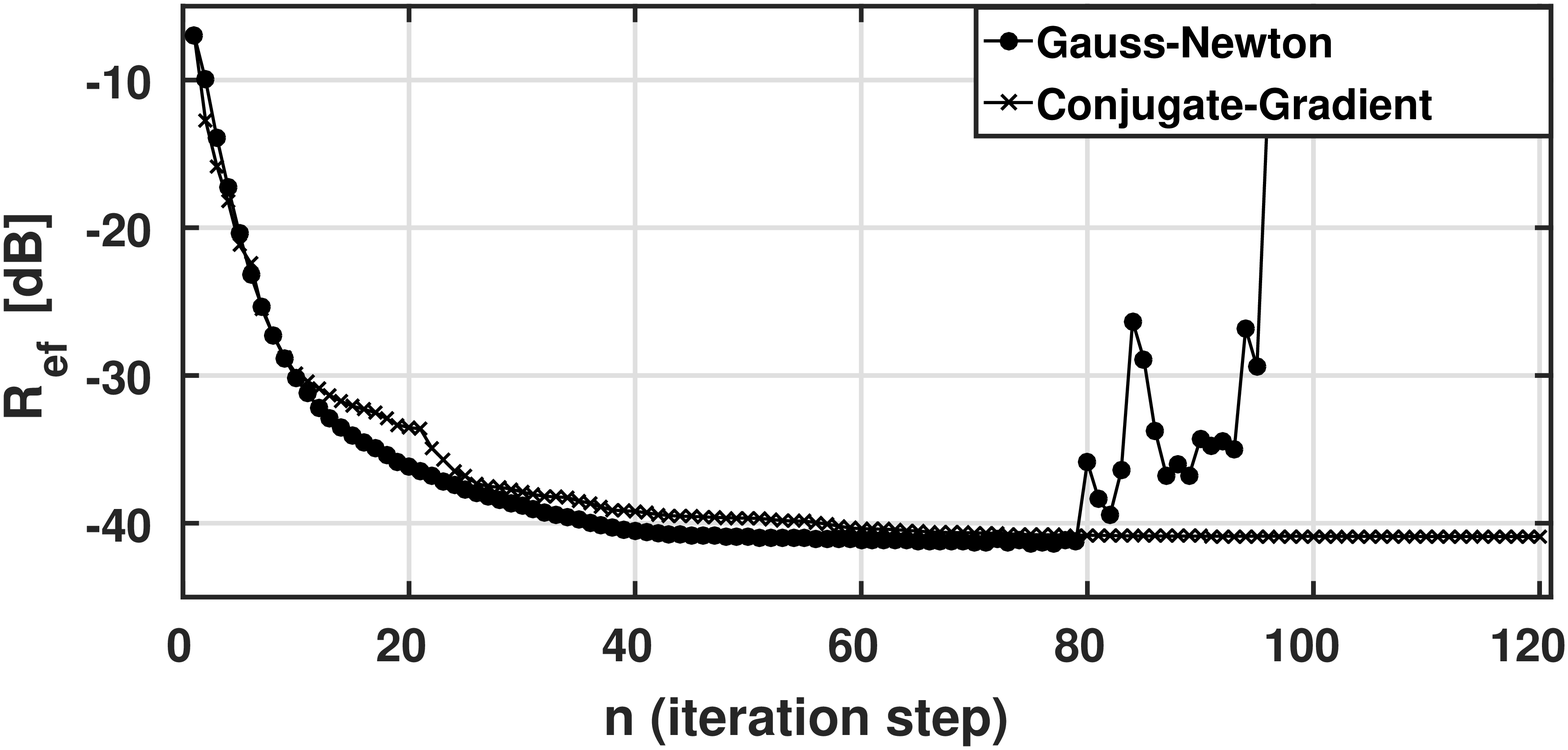} & & & \includegraphics[height=5.5cm,width=8.0cm]{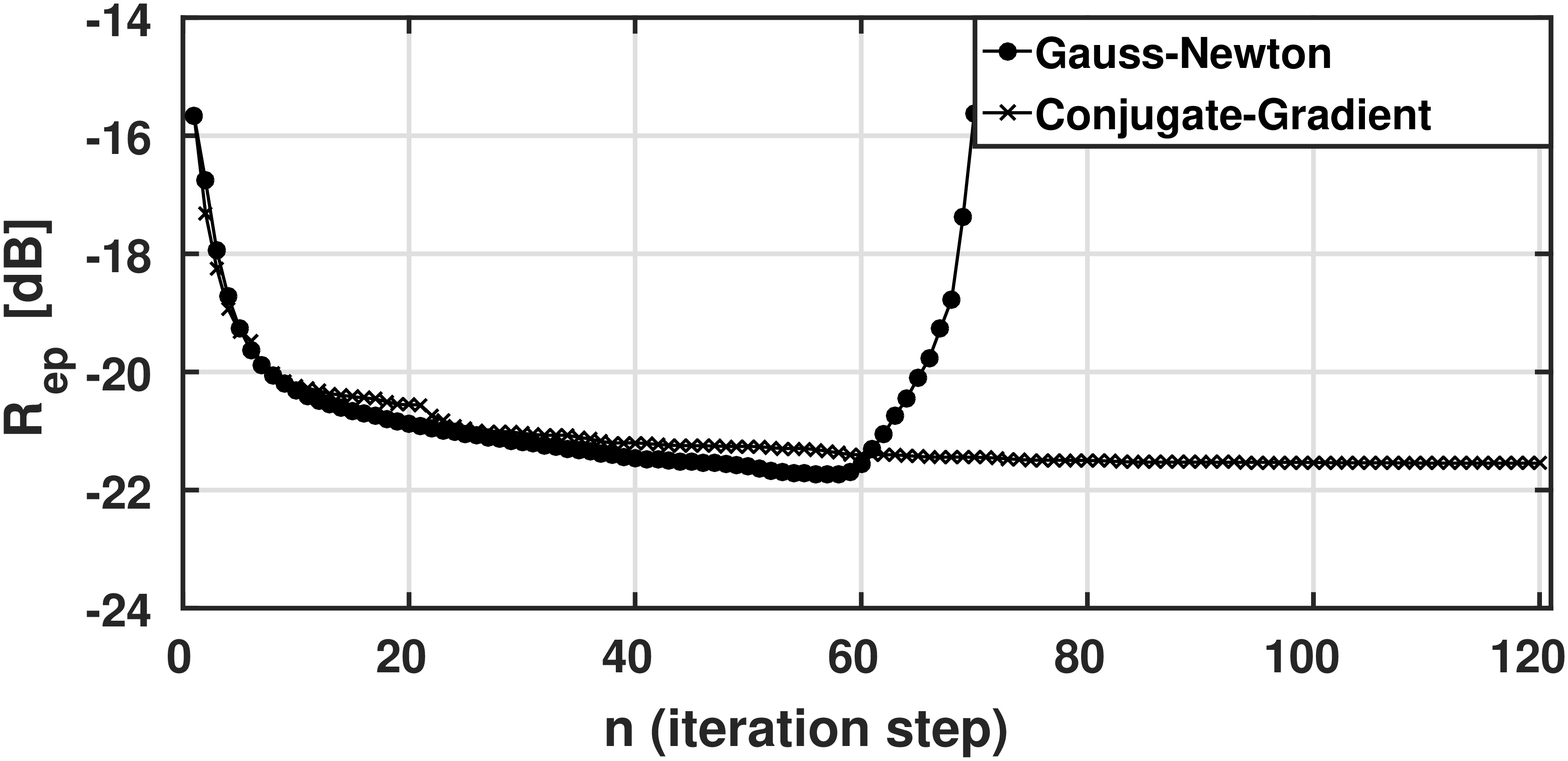} \\
(a) & & & (b) \\
\end{tabular}
\caption{The reference data are noisy with a signal-to-noise ratio ${\rm SNR}=40$. (a): Behaviour of residual errors on the data, calculated in decibels $[{\rm dB}]$ for regularized Gauss-Newton and Conjugate-Gradient reconstructions, according to the number of iterations. (b): Behaviour of residual errors on the complex permittivity-contrast, calculated in decibels $[{\rm dB}]$ for regularized Gauss-Newton and Conjugate-Gradient reconstructions, according to the number of iterations.  
\label{fig1112}}
\end{figure}

\begin{figure}[h]
\centering
\begin{tabular}{cccc}
\includegraphics[height=5.5cm,width=8.0cm]{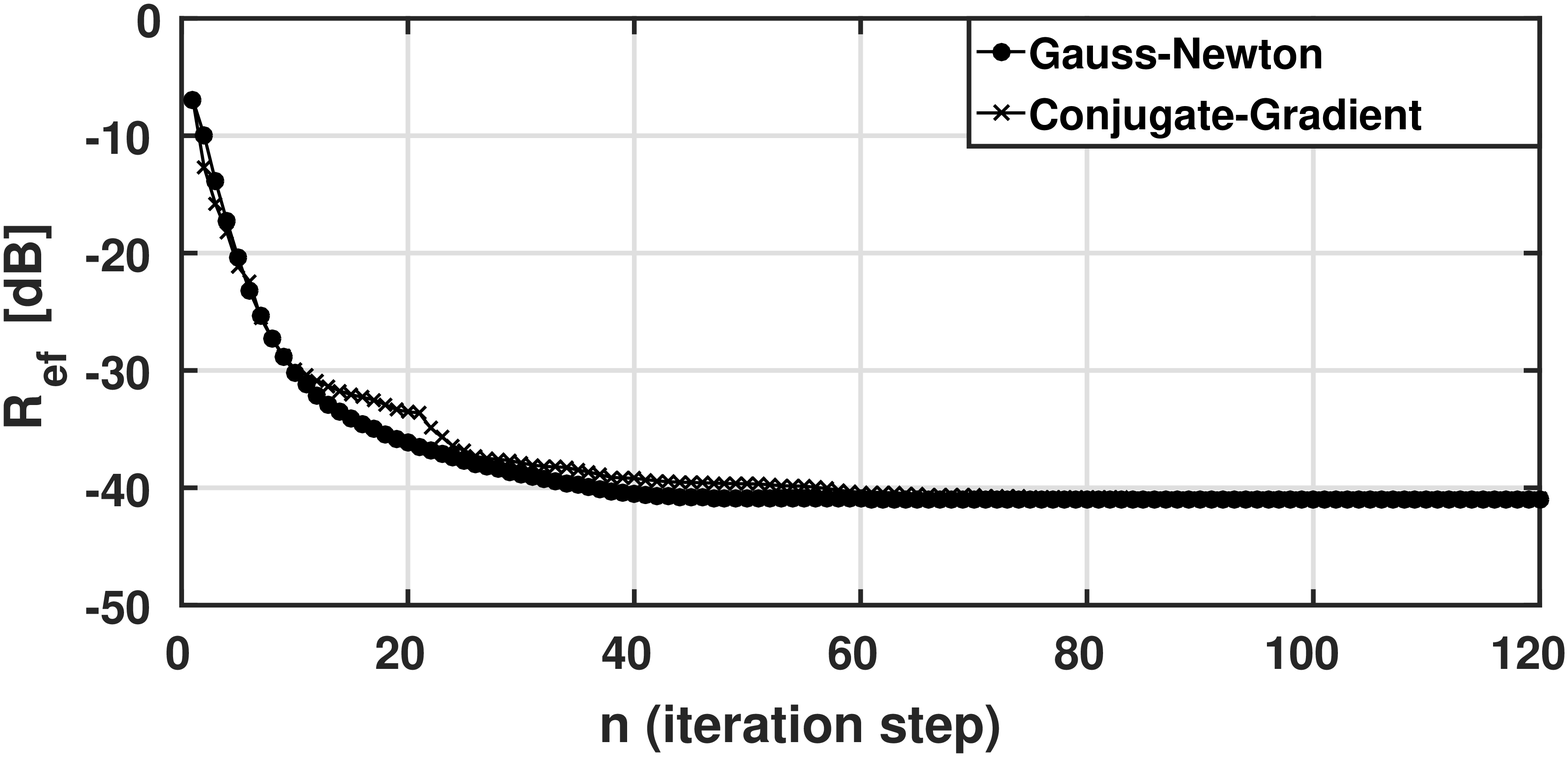} & & & \includegraphics[height=5.5cm,width=8.0cm]{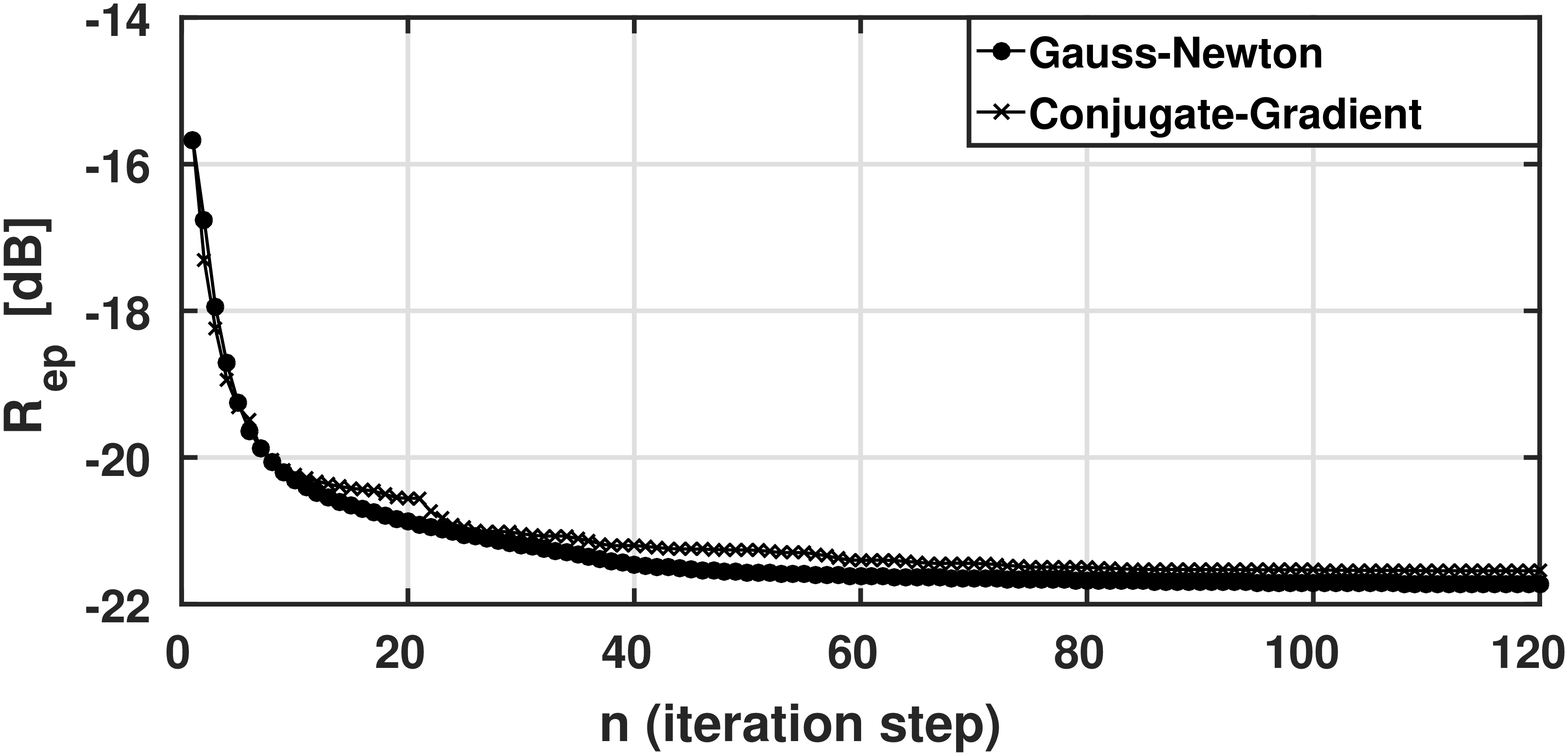} \\
(a) & & & (b) \\
\end{tabular}
\caption{The reference data are noisy with a signal-to-noise ratio ${\rm SNR}=40$. To ensure the convergence of the regularized Gauss-Newton iterative scheme, the value of the Tikhonov parameter at iteration 46 is maintained for all subsequent iterations. (a): Behaviour of residual errors on the data, calculated in decibels $[{\rm dB}]$ for regularized Gauss-Newton and Conjugate-Gradient reconstructions, according to the number of iterations. (b): Behaviour of residual errors on the complex permittivity-contrast, calculated in decibels $[{\rm dB}]$ for regularized Gauss-Newton and Conjugate-Gradient reconstructions, according to the number of iterations. 
\label{fig1314}}
\end{figure}

\begin{figure}[h]
\centering
\begin{tabular}{cccc}
\includegraphics[height=6.5cm,width=7cm]{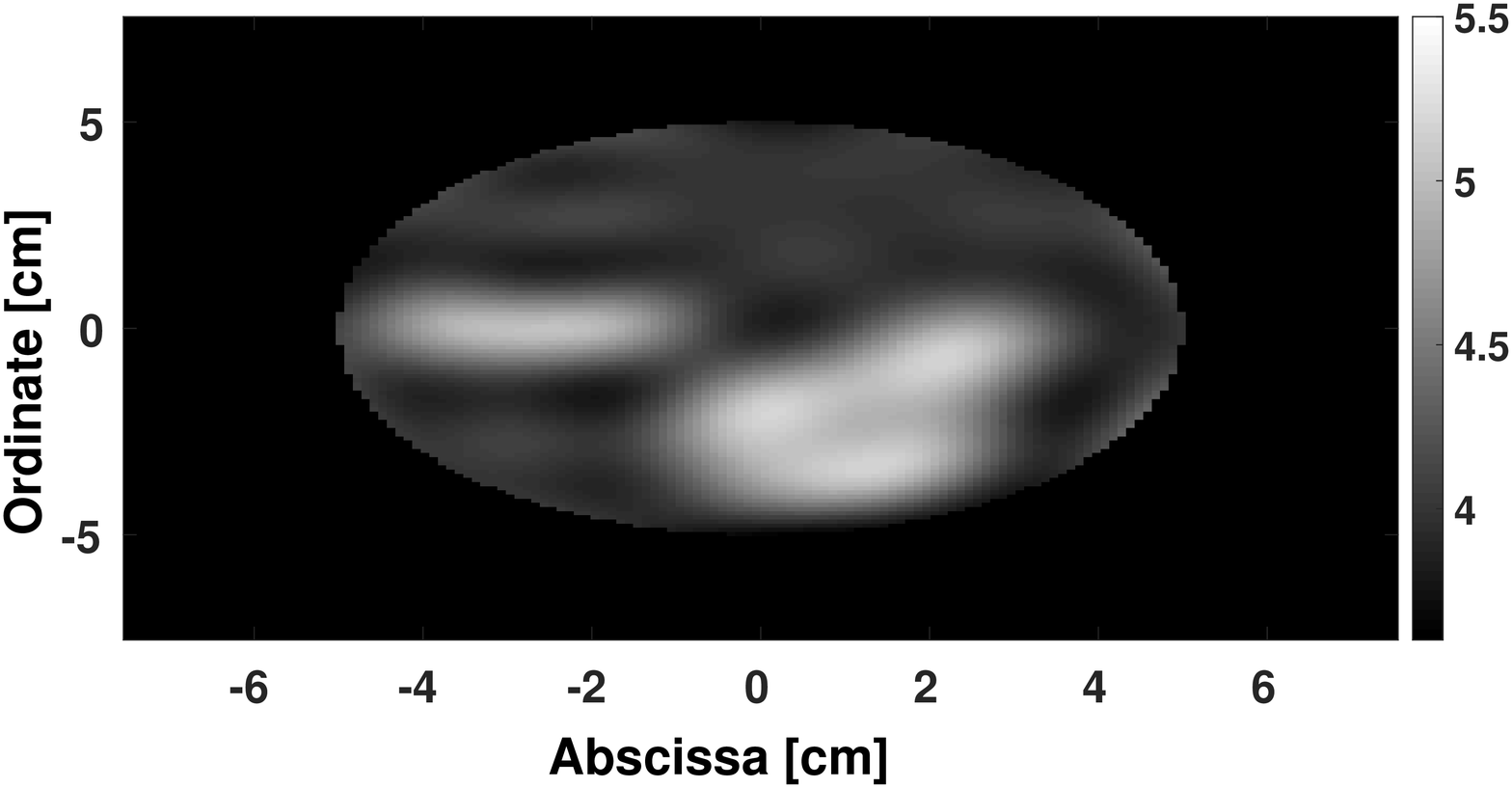} & & & \includegraphics[height=6.5cm,width=7cm]{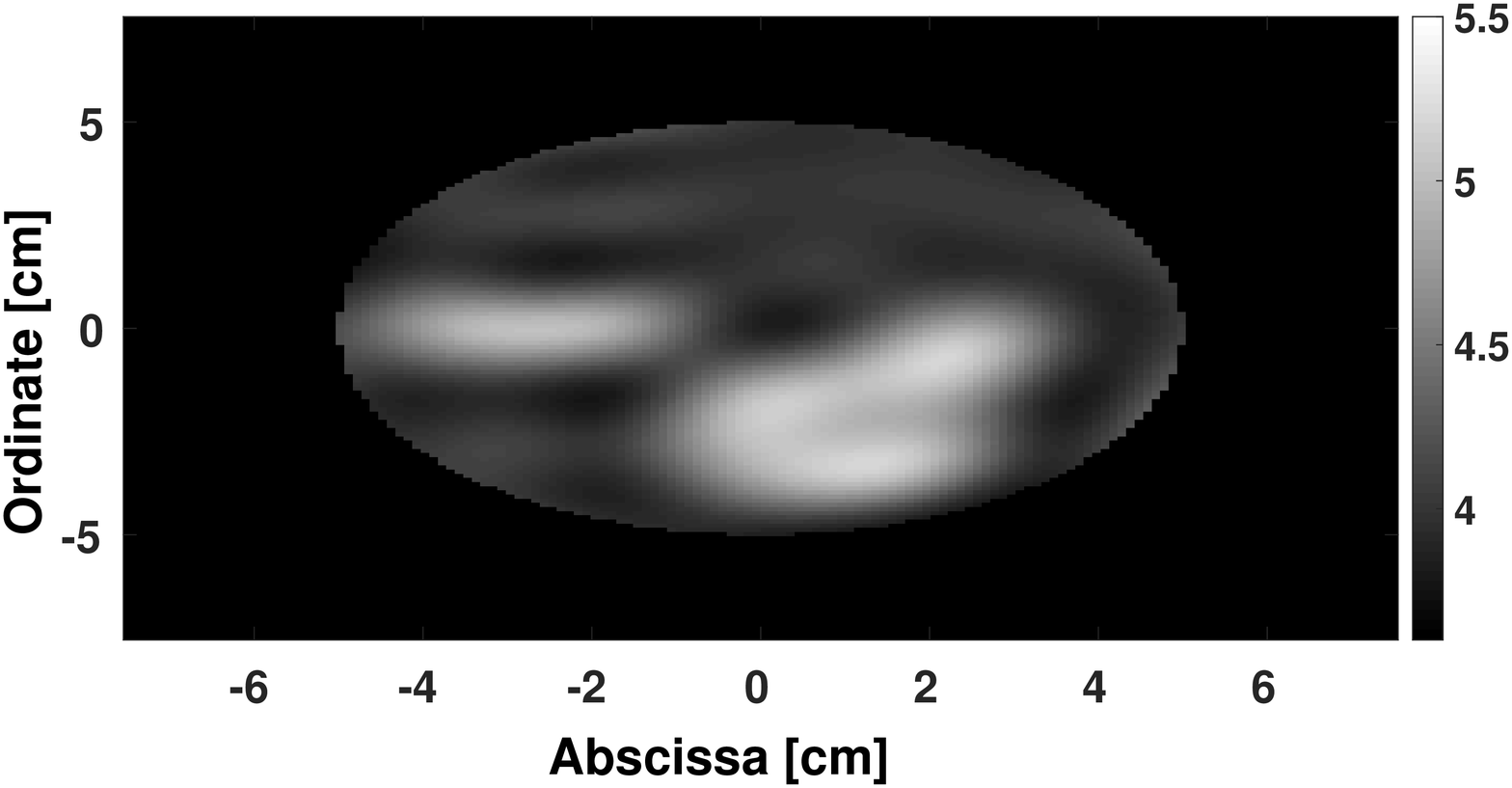} \\
(a) & & & (b) \\
\includegraphics[height=6.5cm,width=7cm]{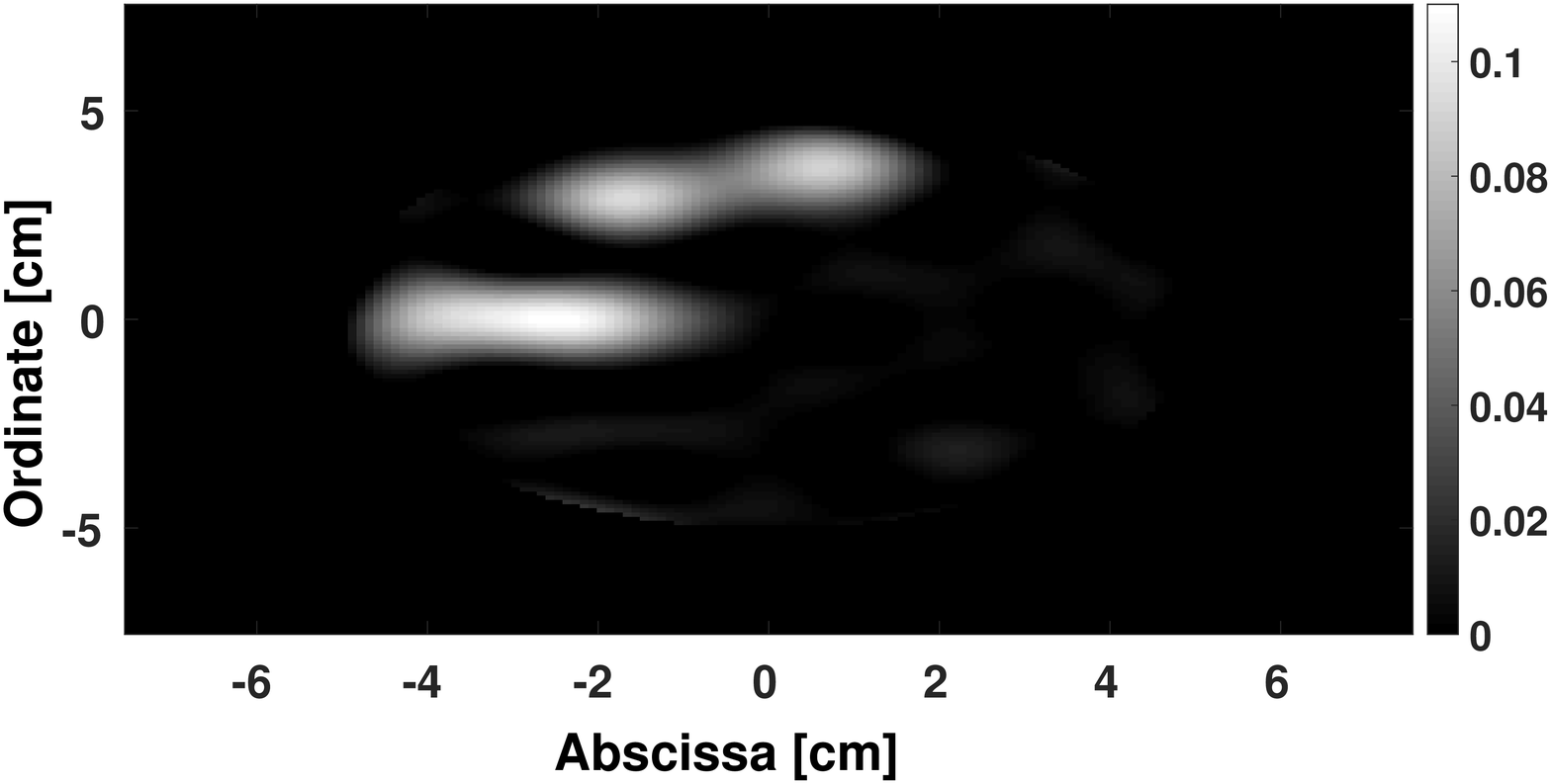} & & & \includegraphics[height=6.5cm,width=7cm]{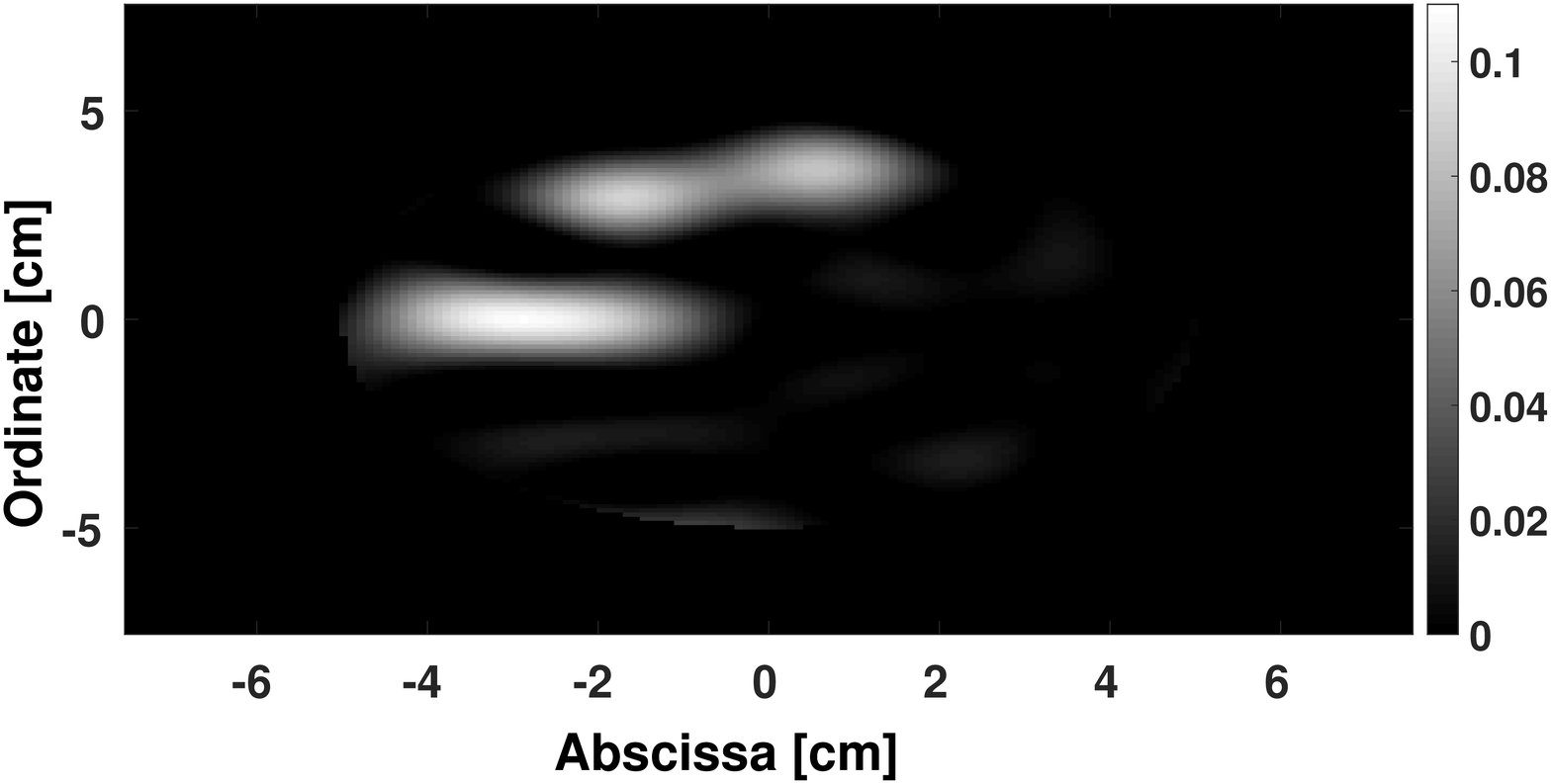} \\
(c) & & & (d) \\
\end{tabular}
\caption{The reference data are noisy with a signal-to-noise ratio ${\rm SNR}=40$. To ensure the convergence of the regularized Gauss-Newton iterative scheme, the value of the Tikhonov parameter at iteration 46 is maintained for all subsequent iterations. (a): relative permittivity $\varepsilon_{_{{{\rm r}}}}^{\widetilde{\protect\scalebox{0.55}{\rm GN}}}(n=120)$ reconstructed after 120 iterations with the regularized Gauss-Newton. (b): relative permittivity $\varepsilon_{_{{{\rm r}}}}^{\protect\scalebox{0.55}{\rm CG}}(n=120)$ reconstructed after 120 iterations with the Conjugate-Gradient. (c): conductivity $\sigma^{\widetilde{\protect\scalebox{0.55}{\rm GN}}}(n=120)$ $({\rm S.m^{\protect\scalebox{0.55}{-1}}})$ reconstructed after 120 iterations with the regularized Gauss-Newton. (d): conductivity $\sigma^{\protect\scalebox{0.55}{\rm CG}}(n=120)$ $({\rm S.m^{\protect\scalebox{0.55}{-1}}})$ reconstructed after 120 iterations with the Conjugate-Gradient.
\label{fig15161718}}
\end{figure}

When we add Gaussian noise to the reference data, with a signal-to-noise ratio of ${\rm SNR}=40$, the residual errors on the field and on the complex permittivity-contrast diverge for the regularized Gauss-Newton iterative scheme, from iteration $80$ for the ${\rm R}_{_{\rm ef}}$ criterion and from iteration $60$ for the ${\rm R}_{_{\rm ep}}$ criterion (see figures \ref{fig1112}-(a) and \ref{fig1112}-(b)), on the other hand, the curves describing these criteria for the Conjugate-Gradient iterative scheme converge towards a plateau (see the figures \ref{fig1112}-(a) and \ref{fig1112}-(b)). To ensure convergence of the regularized Gauss-Newton iterative scheme when the reference data are noisy with a ${\rm SNR}=40$, the value of the Tikhonov parameter of the iteration $46$ is used again for all subsequent iterations, in this case, the residual errors $({\rm R}_{_{\rm ef}},\,{\rm R}_{_{\rm ep}})$ converge well towards a plateau as shown on the figures \ref{fig1314}-(a) and \ref{fig1314}-(b). After performing several tests, we chose to fix the Tikhonov parameter from the iteration $46$, because this value provides the best possible reconstruction and guarantees the convergence of the iterative scheme. The solutions reconstructed by the two iterative approaches are in this case very close to each other, on the figures \ref{fig15161718}-(a), \ref{fig15161718}-(b), \ref{fig15161718}-(c) and \ref{fig15161718}-(d) we find that the reconstruction by the Conjugate-Gradient iterative scheme is identical to that obtained on the noiseless data, the reconstruction by the regularized Gauss-Newton iterative scheme exhibits a lower level of resolution than that of a reconstruction where noiseless data are inverted. \\

\begin{figure}[h]
\centering
\begin{tabular}{cccc}
\includegraphics[height=5.5cm,width=8.0cm]{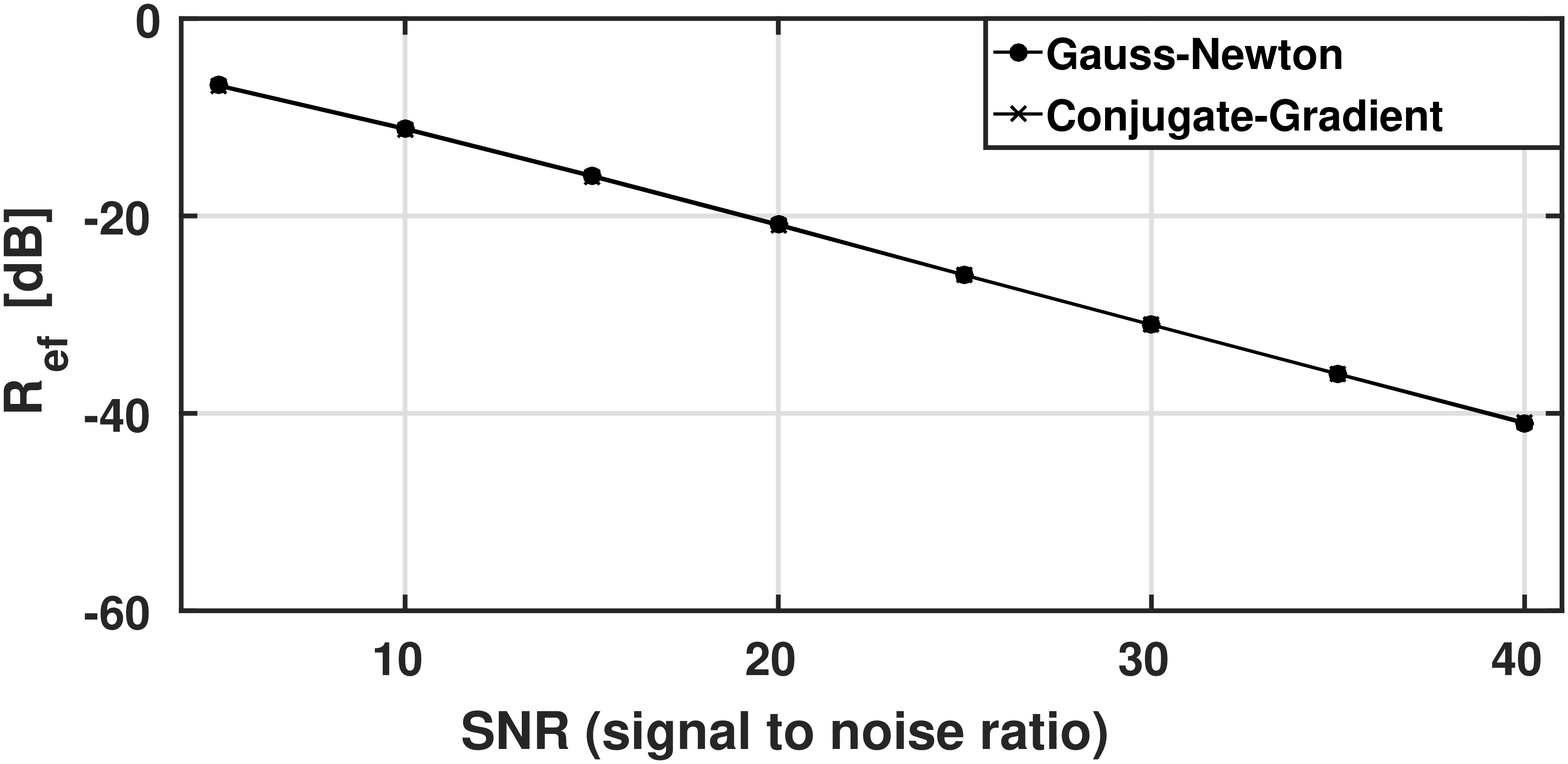} & & & \includegraphics[height=5.5cm,width=8.0cm]{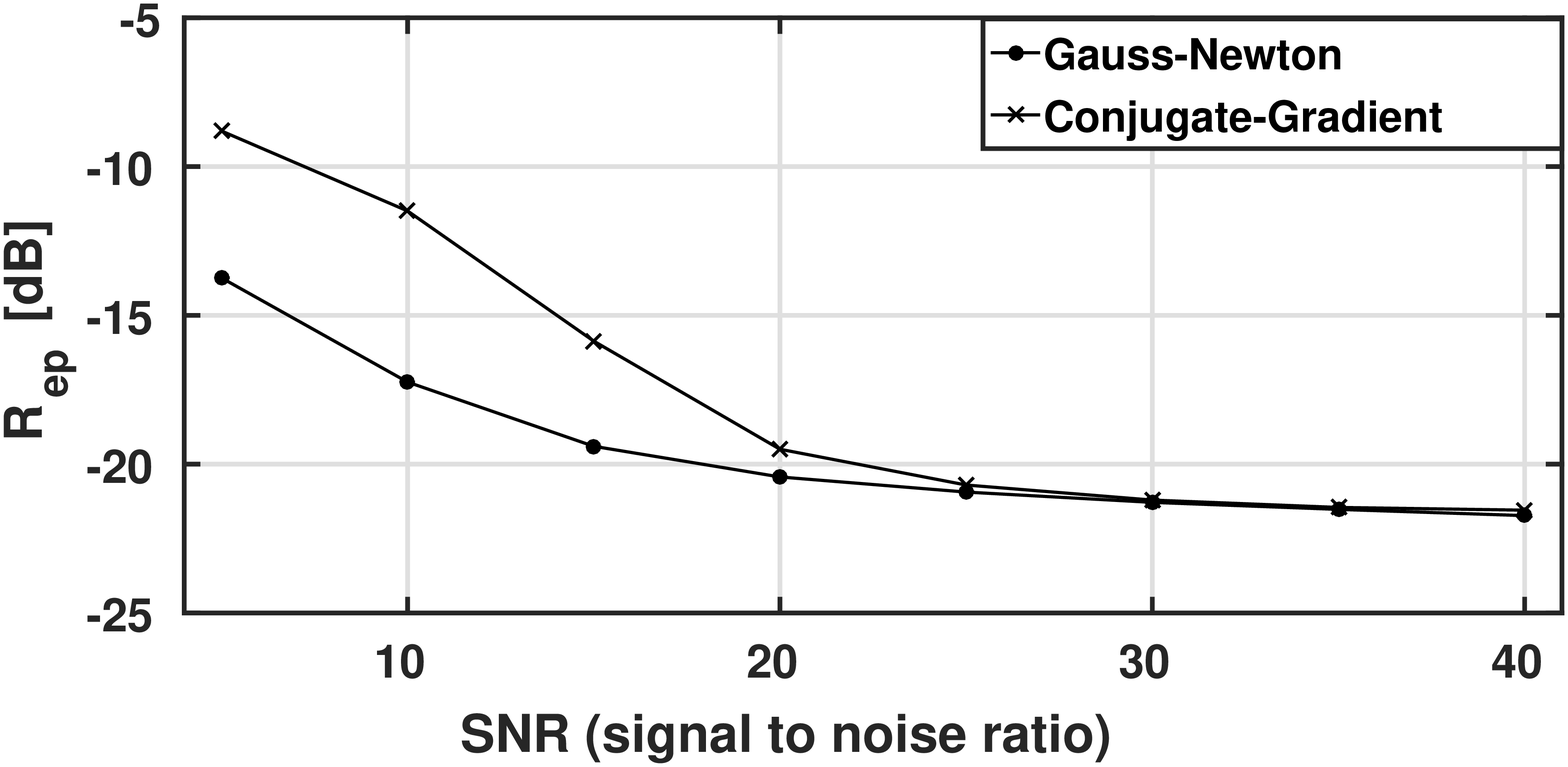} \\
(a) & & & (b) \\
\end{tabular}
\caption{To ensure the convergence of the regularized Gauss-Newton iterative scheme, the value of the Tikhonov parameter at iteration $p$ is maintained for all subsequent iterations, with: $p=46$ for ${\rm SNR}=40$, $p=36$ for ${\rm SNR}=35$, $p=24$ for ${\rm SNR}=30$, $p=9$ for ${\rm SNR}=25$, $p=1$ for ${\rm SNR}=20$, $p=1$ for ${\rm SNR}=15$, $p=1$ for ${\rm SNR}=10$ and $p=1$ for ${\rm SNR}=5$. (a): Behaviour of residual errors on the data, calculated in decibels $[{\rm dB}]$ after $120$ iterations for regularized Gauss-Newton and Conjugate-Gradient reconstructions, according to different values of the signal-to-noise ration $({\rm SNR})$. (b): Behaviour of residual errors on the complex permittivity-contrast, calculated in decibels $[{\rm dB}]$ after $120$ iterations for regularized Gauss-Newton and Conjugate-Gradient reconstructions, according to different values of the signal-to-noise ration $({\rm SNR})$.
\label{fig1920}}
\end{figure}

\begin{figure}[h]
\centering
\begin{tabular}{cccc}
\includegraphics[height=6.5cm,width=7cm]{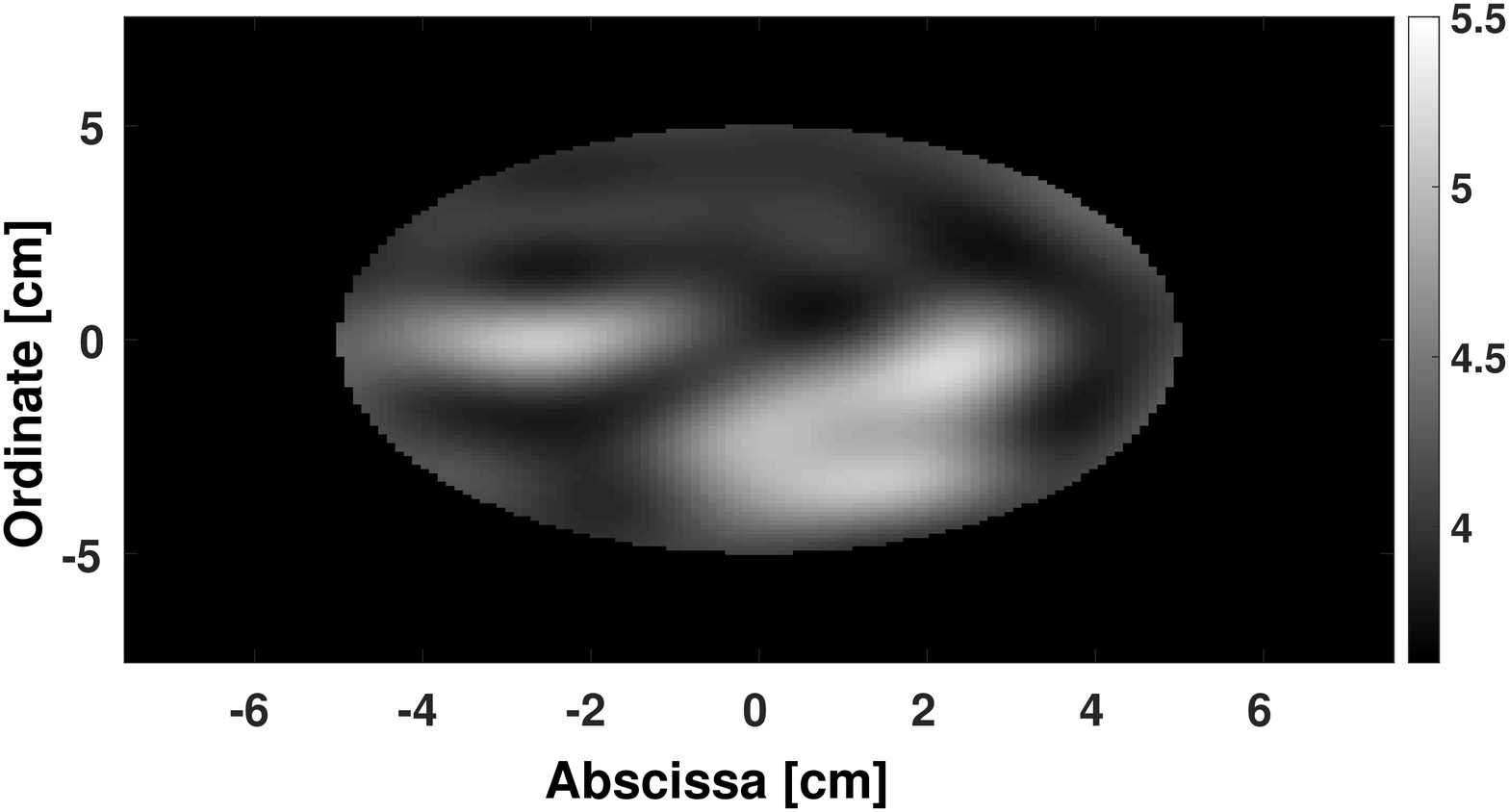} & & & \includegraphics[height=6.5cm,width=7cm]{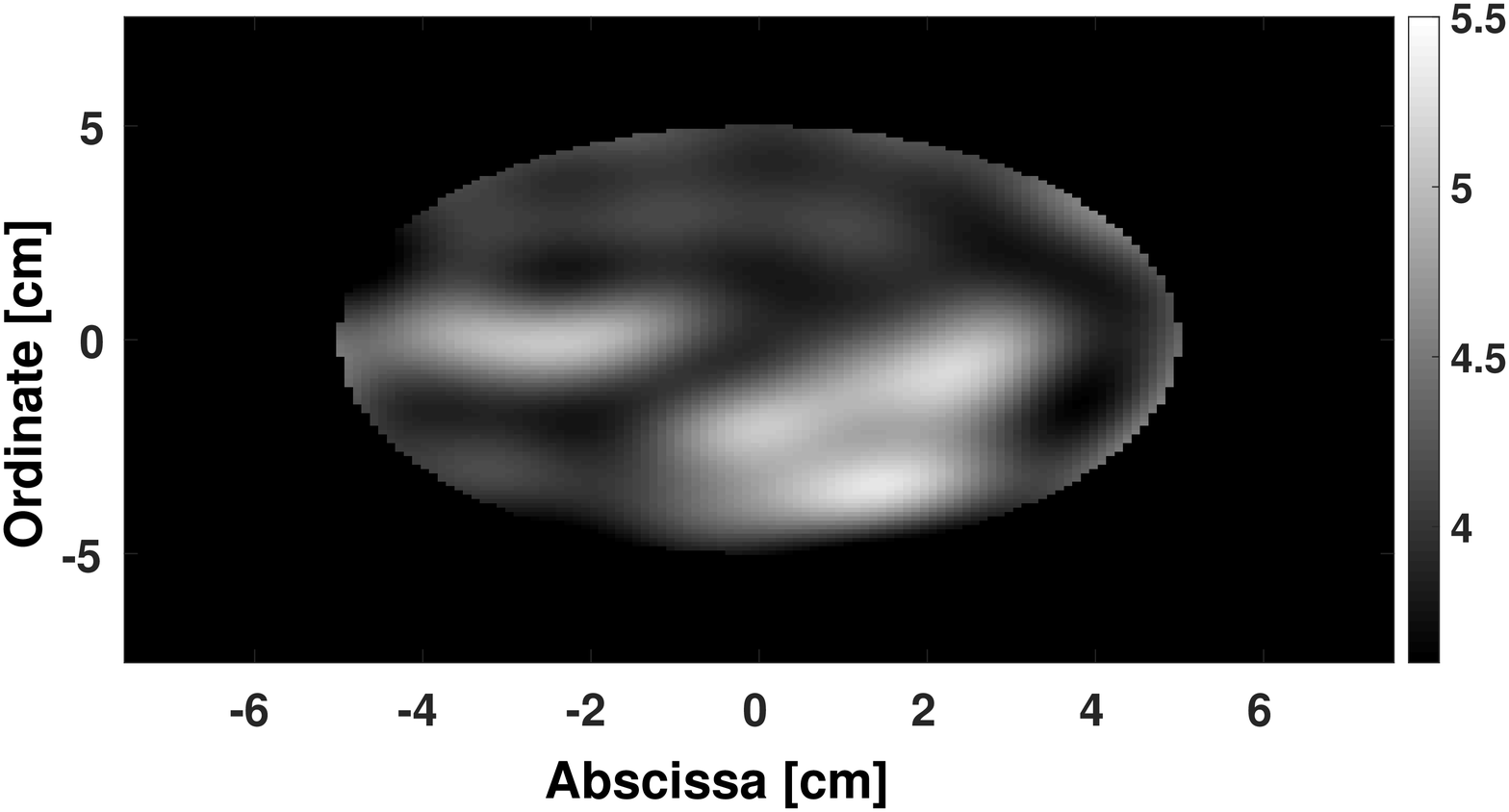} \\
(a) & & & (b) \\
\includegraphics[height=6.5cm,width=7cm]{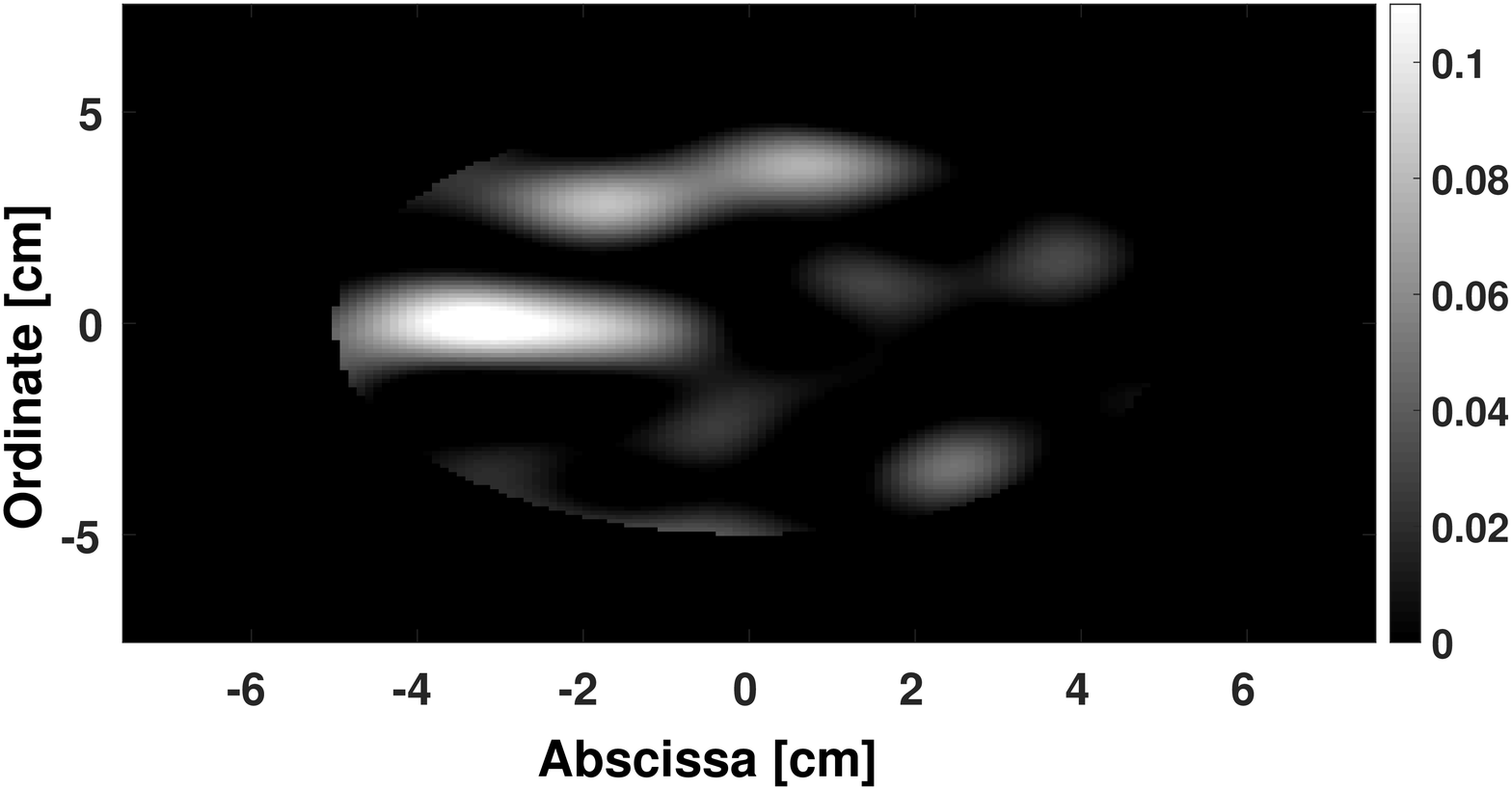} & & & \includegraphics[height=6.5cm,width=7cm]{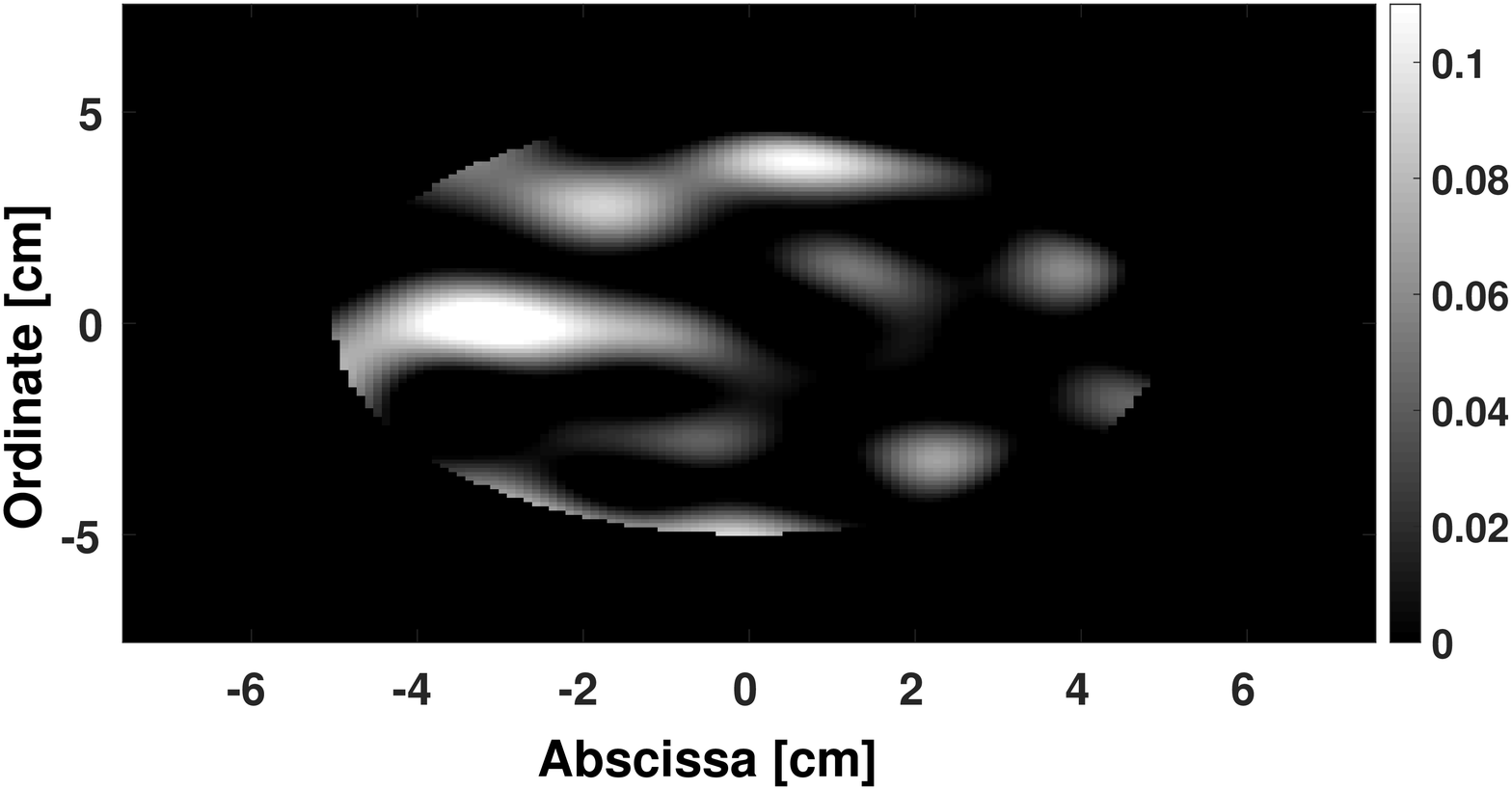} \\
(c) & & & (d) \\
\end{tabular}
\caption{The reference data are noisy with a signal-to-noise ratio ${\rm SNR}=20$. To ensure the convergence of the regularized Gauss-Newton iterative scheme, the value of the Tikhonov parameter at iteration $1$ is maintained for all subsequent iterations. (a): relative permittivity $\varepsilon_{_{{{\rm r}}}}^{\widetilde{\protect\scalebox{0.55}{\rm GN}}}(n=120)$ reconstructed after $120$ iterations with the regularized Gauss-Newton. (b): relative permittivity $\varepsilon_{_{{{\rm r}}}}^{\protect\scalebox{0.55}{\rm CG}}(n=120)$ reconstructed after $120$ iterations with the Conjugate-Gradient. (c): conductivity $\sigma^{\widetilde{\protect\scalebox{0.55}{\rm GN}}}(n=120)$ $({\rm S.m^{\protect\scalebox{0.55}{-1}}})$ reconstructed after $120$ iterations with the regularized Gauss-Newton. (d): conductivity $\sigma^{\protect\scalebox{0.55}{\rm CG}}(n=120)$ $({\rm S.m^{\protect\scalebox{0.55}{-1}}})$ reconstructed after $120$ iterations with the Conjugate-Gradient.
\label{fig21222324}}
\end{figure}

The figures \ref{fig1920}-(a) and \ref{fig1920}-(b) show the evolution of the residual errors $({\rm R}_{_{\rm ef}},\,{\rm R}_{_{\rm ep}})$ obtained at iteration $120$, for different values of the signal to noise ratio. Concerning the regularized Gauss-Newton iterative scheme, for each value of ${\rm SNR}$, we perform several tests to find the optimal iteration from which the Tikhonov parameter will be fixed, this iteration is optimal because it guarantees both the convergence of the iterative process and the reconstruction of the object with the most possible resolution. On the figures \ref{fig1920}-(a) and \ref{fig1920}-(b), we see that the residual error on the field ${\rm R}_{_{\rm ef}}$ evolves in the same way for both approaches. The residual error on the complex permittivity-contrast ${\rm R}_{_{\rm ep}}$ takes for both iterative approaches the same values when we decrease the ${\rm SNR}$, every $5$ step, from the value $40$ to the value $25$, below, we find that this error increases more significantly for the Conjugate-Gradient iterative scheme. For instance, on the figures \ref{fig21222324}-(a), \ref{fig21222324}-(b), \ref{fig21222324}-(c) and \ref{fig21222324}-(d) obtained for a signal-to-noise ratio ${\rm SNR}=$20, we find that the solution of the Conjugate-Gradient iterative scheme is further deteriorated compared to that of the regularized Gauss-Newton iterative scheme. A comparison between the reconstructions obtained by the two iterative schemes for a lower value of ${\rm SNR}$ shows that it is difficult to affirm that the regularized Gauss-Newton iterative scheme conserves this advantage, even if objectively its residual error ${\rm R}_{_{\rm ep}}$ remains lower than that of the Conjugate-Gradient.

\section{Conclusion}
\label{sec:Conclusion}
This work provided an opportunity to numerically compare the regularized Gauss-Newton and Conjugate-Gradient iterative schemes in the context of the reconstruction of relative permittivity and conductivity in microwave tomography. On the inversion of noiseless data, the regularized Gauss-Newton iterative scheme with an adjustable regularization parameter during the iterations, allowed to reconstruct the object with a better resolution than that of a reconstruction by the Conjugate-Gradient iterative scheme. For noisy data with a reasonable signal-to-noise ratio, both approaches give almost identical reconstructions. We conclude that to invert experimental data, it is more practical to use the Conjugate-Gradient iterative scheme. It is robust as long as signal-to-noise ratios remain reasonable, which is often the case in the microwave imaging community where the measurements produced are increasingly controlled. The Conjugate-Gradient iterative scheme offers many advantages, it is fast in terms of computation time per iteration, with its scale parameter that is analytically calculable from the Gauss-Newton equation, it becomes completely autonomous. This last point is very important, because by avoiding the control of the inversion process by any external fine tuning, we make it accessible to users who are not specialized in inverse problem solving.

\appendixx{Calculus of gradient and Hessian}
\label{gradhess}

To deduce the quadratic model $\mathcal{M}^{^{\scalebox{0.55}{\rm GN}}}(\,\zeta\,)$, let's start by calculating $\mathcal{F_{_C}}(\,\chi\,+\,\zeta\,)\,-\,\mathcal{F_{_C}}(\,\chi\,)$ : \\
\begin{equation}\label{diffcost}
\begin{array}{l}
\mathcal{F_{_C}}(\,\chi\,+\,\zeta\,)\,-\,\mathcal{F_{_C}}(\,\chi\,)\,=\, \\
\\
\frac{1}{\sum\limits_{lm=1}^{M^{\scalebox{0.55}{2}}-M}|{\rm E}^{^{\scalebox{0.7}{\rm s,ex}}}_{_{lm}}|^{^{\scalebox{0.55}{2}}}}\,\times\,\sum\limits_{lm=1}^{M^{\scalebox{0.55}{2}}-M}(\,[{\rm E}^{^{\scalebox{0.7}{\rm s,ex}}}_{lm}\,-\,{\rm E}^{^{\scalebox{0.7}{\rm s}}}_{_{lm}}(\,\chi\,+\,\zeta\,)]\,\,[{\rm E}^{^{\scalebox{0.7}{\rm s,ex}}}_{lm}\,-\,{\rm E}^{^{\scalebox{0.7}{\rm s}}}_{_{lm}}(\,\chi\,+\,\zeta\,)]^{^{*}}\,-\,[{\rm E}^{^{\scalebox{0.7}{\rm s,ex}}}_{lm}\,-\,{\rm E}^{^{\scalebox{0.7}{\rm s}}}_{_{lm}}(\,\chi\,)]\,\,[{\rm E}^{^{\scalebox{0.7}{\rm s,ex}}}_{lm}\,-\,{\rm E}^{^{\scalebox{0.7}{\rm s}}}_{_{lm}}(\,\chi\,)]^{^{*}}),
\end{array}
\end{equation}

With the symbol $*$ that designates the complex conjugate. By introducing the functional expansion limited to its first order ${\rm E}^{^{\scalebox{0.7}{\rm s}}}_{_{lm}}(\,\chi\,+\,\zeta\,)\,\simeq\,{\rm E}^{^{\scalebox{0.7}{\rm s}}}_{_{lm}}(\,\chi\,)\,+\,\zeta^{^{\scalebox{0.55}{\rm T}}}\,{\rm g}_{_{lm}}$, where the symbol ${\rm T}$ designates the transpose and ${\rm g}_{_{lm}}\,=\,-{\rm i}\,\omega\,\varepsilon_{\scalebox{0.55}{0}}\,\Delta\mathcal{V}\,\widehat{\rm E}_{_l}\,{\rm E}_{_m}$ denotes the gradient of the datum ${\rm E}^{^{\scalebox{0.7}{\rm s}}}_{_{lm}}(\,\chi\,)$ with respect to the complex permittivity-contrast $\chi$, calculated with the reciprocity gap functional method \cite{arhab2018high,roger1982reciprocity}. Here $\widehat{\rm E}_{_l}$ denotes a diagonal matrix of dimension $Q \times Q$, and whose diagonal elements are the elements of the vector ${\rm E}_{_l}$ of dimension $Q \times 1$, the vector ${\rm E}_{_m}$ of dimension $Q \times 1$ has as elements the values of the field solution of the state equation \ref{statequation}, of incident field ${\rm E}^{^{\scalebox{0.7}{\rm i}}}_{_m}({\mb r})\,=\,-\frac{1}{4}\,\omega\,\mu_{\scalebox{0.55}{0}}\,\,{\rm H}^{^{{\scalebox{0.55}{(1)}}}}_{\scalebox{0.55}{0}}({\rm k}\,|{\mb r}-{\mb r}_{_m}|)$. The above expression can after some calculation steps be approached to give the following Gauss-Newton quadratic model $\mathcal{M}^{^{\scalebox{0.55}{\rm GN}}}(\,\zeta\,)$: \\

\begin{equation}\label{diffcost}
\begin{array}{l}
\mathcal{F_{_C}}(\,\chi\,+\,\zeta\,)\,-\,\mathcal{F_{_C}}(\,\chi\,)\,\simeq\,\mathcal{M}^{^{\scalebox{0.55}{\rm GN}}}(\,\zeta\,)\,= \\
\\
\Re\{\zeta^{^{\scalebox{0.55}{\dag}}}\,\frac{-2}{\sum\limits_{lm=1}^{M^{\scalebox{0.55}{2}}-M}|{\rm E}^{^{\scalebox{0.7}{\rm s,ex}}}_{_{lm}}|^{^{\scalebox{0.55}{2}}}}\,\sum\limits_{lm=1}^{M^{\scalebox{0.55}{2}}-M}{\rm g}_{_{lm}}^{^{*}}\,(\,{\rm E}^{^{\scalebox{0.7}{\rm s,ex}}}_{lm}\,-\,{\rm E}^{^{\scalebox{0.7}{\rm s}}}_{_{lm}}(\,\chi\,)\,)\}\,+\,\frac{1}{2}\,\zeta^{^{\scalebox{0.55}{\dag}}}\,\frac{+2}{\sum\limits_{lm=1}^{M^{\scalebox{0.55}{2}}-M}|{\rm E}^{^{\scalebox{0.7}{\rm s,ex}}}_{_{lm}}|^{^{\scalebox{0.55}{2}}}}\,\sum\limits_{lm=1}^{M^{\scalebox{0.55}{2}}-M}{\rm g}_{_{lm}}^{^{*}}\,{\rm g}_{_{lm}}^{\scalebox{0.55}{\rm T}}\,\zeta
\end{array}
\end{equation}

Therefore, we find a quadratic model of the form $\mathcal{M}^{^{\scalebox{0.55}{\rm GN}}}(\,\zeta\,)\,=\,\Re\{\zeta^{^{\scalebox{0.55}{\dag}}}\,\mathcal{G}\}\,+\,\frac{1}{2}\,\zeta^{^{\scalebox{0.55}{\dag}}}\,\mathcal{H}^{^{\scalebox{0.55}{\rm GN}}}\,\zeta$, with: \\

\begin{equation}\label{gradhessfound}
\left\lbrace\begin{array}{l}
\mathcal{G}\,=\,\frac{-2}{\sum\limits_{lm=1}^{M^{\scalebox{0.55}{2}}-M}|{\rm E}^{^{\scalebox{0.7}{\rm s,ex}}}_{_{lm}}|^{^{\scalebox{0.55}{2}}}}\,\sum\limits_{lm=1}^{M^{\scalebox{0.55}{2}}-M}{\rm g}_{_{lm}}^{^{*}}\,(\,{\rm E}^{^{\scalebox{0.7}{\rm s,ex}}}_{lm}\,-\,{\rm E}^{^{\scalebox{0.7}{\rm s}}}_{_{lm}}(\,\chi\,)\,) \\
\\
\mathcal{H}^{^{\scalebox{0.55}{\rm GN}}}\,=\,\frac{+2}{\sum\limits_{lm=1}^{M^{\scalebox{0.55}{2}}-M}|{\rm E}^{^{\scalebox{0.7}{\rm s,ex}}}_{_{lm}}|^{^{\scalebox{0.55}{2}}}}\,\sum\limits_{lm=1}^{M^{\scalebox{0.55}{2}}-M}{\rm g}_{_{lm}}^{^{*}}\,{\rm g}_{_{lm}}^{\scalebox{0.55}{\rm T}}
\end{array}\right.
\end{equation}

\end{document}